\documentclass{elsart}
\usepackage{graphics}
\begin{document}
\begin{frontmatter}

\title{The Canonical Heat Capacity of Normal Mesoscopic Fermion Systems: The Temperature Evolution and Particle Number Oscillations}
\author[Radium]{N.K. Kuzmenko\corauthref{cor}},
\corauth[cor]{Corresponding author.}
\ead{Kuzmenko@NK9433.spb.edu}
\author[Univ]{V.M. Mikhajlov}
\address[Radium]{V.G.Khlopin Radium Institute, 194021
St.-Petersburg, Russia}
\address[Univ]{Institute of Physics St.--Petersburg State
University 198904, Russia}

\date{\today}
\begin{abstract}
The heat capacity of an individual mesoscopic normal (nonsuperconducting) fermion system treated as a canonical ensemble of independent particles confined in a spatial area with fixed confinement parameters is studied in a wide range of particle numbers $N<10^5$ and temperatures which vary from values close to zero up to the Fermi energy $\varepsilon_F$. The temperature evolution of the heat capacity is naturally divided into four stages. On the first one the heat capacity exponentially increases with temperature and at a resonance temperature reaches either a local maximum or an irregularity in its growth. This temperature being measured can give information concerning the level spacings $\Delta\varepsilon$ in the immediate proximity of $\varepsilon_F$. Calculations of $\Delta\varepsilon$ ( as the difference between the Fermi level and the level just above it) performed for diverse systems confined within nonsymmetric oscillator or rectangular wells show that nearly uniform distributions $\Delta\varepsilon$ is stretched up to $\sim(2\div 3)\delta_F$ ($\delta_F$ is the averaged level spacing in the vicinity of the Fermi energy). On the second stage ($T$ is more or of order of $\Delta\varepsilon$) the progressive suppression of the level density oscillations takes place. During this stage the heat capacity oscillations v.s. particle number can be distinctly observed. These oscillations give the particle number variation of the temperature averaged level density. The $N$-oscillations discontinue for all $3D$-systems at the temperature of order of $\varepsilon_F/N^{1/3}$ and for $2D$-systems at $\varepsilon_F/N^{1/2}$. The growth of the heat capacity on the third stage of the evolution is governed by the $T$-linear law if the particle number is large enough ($N>10^3$). In small systems ($N<10^3$) the heat capacity reveals marked deviations from this law immediately near the temperature suppressing the level density oscillations. It is found for rectangular potentials that the Sommerfeld factor $\alpha$ in the linear law  ($C=\alpha T$) for such small systems shows more complicated $N$-dependence as compared with large systems where $\alpha$ is strictly proportional to $N$. For temperatures tending to $\varepsilon_F$ the deviations from the $T$ linearity are evident and at $T>\varepsilon_F$ (the fourth stage of the evolution) any system attains to the classic Boltzmann-Maxwell limit irrespective of the particle number. All our calculations are carried out by using the canonical polynomial method. The results of approximate methods  (in which the Fermi-Dirac function is used as the occupation probability and the chemical potential $\lambda$ is temperature dependent to allow for the particle number conservation on the average) are shown for some exemplary cases which indicate that the approximation explicitly taking into account  $\partial\lambda/\partial T$ is almost equivalent to the canonical method excepting a temperature region $T<\delta_F$.
\end{abstract}

\end{frontmatter}

\section{Introduction}
The first investigation of the mesoscopic fermion heat capacity belongs apparently to Fr\"{o}lich , Ref.~\cite{frolich} of $1937$, who considered a system with a single electron spectrum similar to the spherical oscillator, i.e. with equal level spacing $\Delta\varepsilon$ and high level degenerations that was regarded as an approximation to the spectrum of a cubic rectangular potential. This model allowed Fr\"{o}lich to establish such distinctive feature of the low temperature heat capacity ($C$) as exponential growth with increasing temperature. He also found that at subsequent heating up to $k_BT\sim 2\Delta\varepsilon$ the specific heat gains the bulk value.

The most important objection against Fr\"{o}lich's approach was given by Kubo (summary of the investigations of Kubo and his collaborators is given in Refs~\cite{perenboom,halperin}) who emphasized the irregularities of level spacings in mesoscopic systems which, as it seemed untill the eighties of the past century, could not have a defined and fixed shape and size. Therefore for describing an ensemble of such objects Kubo introduced into consideration the probability distribution for the spacings between adjacent levels. In this case the thermodynamic quantities had to be calculated not with the spectrum of an individual system but with interlevel spacings which possess some probability of their appearance. As Kubo supposed the Poisson distribution in which very small spacings are most probable this immediately results in practical disappearance of the exponential part of the heat capacity and even at $T/\delta<1$ ($\delta$ is the averaged level spacing) $C$ becomes linear in temperature. Application of other types of the level statistics remains rapid growth of $C$ v.s. $T$ though now it is not exponential increasing but proportional to either $T^{\,2}$ or $T^{\,4}$ \cite{perenboom}.

However the interest in such description of mesoscopic systems has partly diminished after the discovery of the shell structure in metal clusters~\cite{heer} and appearance of the possibility to fabricate mesoscopic systems such as artificial atoms or quantum dots with given geometry and controlled particle numbers~\cite{kouwen,delft}. Electron-shell effects were also identified in experiments on diverse metal nanowires~\cite{yanson,diaz,mares}. One more stimulus for the consideration of individual fermion systems came from experiments on trapped gases of fermions ~\cite{pethick}.

Therefore on this new stage in development of mesoscopic physics when variety of phenomena could be explained only if the level structure was explicitly taken into account the main interest has been concentrated on investigations of individual systems with a fixed particle number and geometry.

As any of such systems can exchange energy with surroundings it has to be treated as a canonical ensemble. For the first time an analytical formalism (approximate in general but exact for very large particle number) was worked out by Denton, M\"uhlschlegel and Scalpino~\cite{denton} for $1D$-oscillator. Brack et al~\cite{brack} applied the canonical formalism (though not in its simplest version) to calculations of the electron heat capacity of alkali clusters with realistic single electron spectrum created by the temperature self-consistent potential. Parallel with exact canonical calculations approximate methods have been developed: partial projection, Refs.~\cite{lang,frauendorf}, saddle-point approximation, Ref.~\cite{rossignoli}, integral-transformation method, Ref.~\cite{schon}. In this chapter we apply our canonical polynomial formalism, Ref.~\cite{kuzmenko}, the general idea of which is similar to that of Denton et al however our method is applied to any energy level distribution and arbitrary particle numbers.

Though among many experiments performed on mesoscopic systems to the present time there are only several investigations of the thermal behavior and in particular the heat capacity of normal and superconducting mesoscopic systems (see e.g. Ref.~\cite{burgeois} and references therein) we believe that such experiments will gain greater development owing to their practical applications. Therefore preliminary theoretical estimations of $C$ for individual systems in simplified models that we give in this chapter can be a basis for arrangements of future experiments and further theoretical improvement.

Our calculations of the fermion heat capacity are carried out in the independent particle model that can be directly applied only to trapped ultra-cold Fermi-gases. In the majority of the electron systems the strong interaction of each electron with ion surroundings and other electrons forms fermion quasiparticles with an effective mass different from the free electron mass and an energy spectrum similar to that for free electrons in a confinement potential. As shown in the Fermi liquid theory~\cite{abrikosov} the quasiparticle potential ( the existence of which is directly established by experiments on the observation of the shell structure) can depend on the excitation energy and temperature that does not explicitly taken into account in our calculations - this is our first simplification. The second simplification consists in neglecting the quasiparticle damping i.e. levels are considered to be perfectly sharp. However both these simplification can affect only high temperature values of $C$. Besides, we do not allow for the residual electron-phonon interaction i.e. such its part which in an averaged form does not enter into the electron mean field. As known~\cite{halperin} this interaction gives rise to renormalization of Sommerfeld's constant $\alpha$ for bulk metals at those temperatures that lead to the linear variation of $C$ v.s. $T$ ($C=\alpha T$) therefore our estimations of $C$ can be essentially corrected by taking into account the electron-phonon interaction in this temperature region. However as shown below at these temperatures (when $C$ is proportional $T$) the heat capacity practically does not bear information concerning the level structure near the Fermi energy and so we hope that the electron-phonon interaction being allowed for does not change drastically our conclusions.

Two types of confinement are employed in our calculations. As a hard confinement we make use of the $2D$ and $3D$ rectangular potentials while the harmonic oscillator potentials are regarded as a soft confinement. The realistic case is something intermediate between these potentials. This can  reveal itself again in the temperature region that provides the linear variation of $C$ with increasing $T$. The finite depth of the potential can affect the high temperature variations of $C$ which is certainly only of academic interest in the majority of cases.

The content of this chapter includes $7$ sections. In Sec.2 the basic formulae for the calculation of $C$ and interpretation of numerical results are given, here also the stages of the temperature evolution of $C$ are outlined in consecutive order. Then in Sec.3-6 we give the results corresponding to each stage. In Sec.7 we present conclusions.

\section{The heat capacity as an average of the level density}

Thermodynamic properties of fermion isolated mesoscopic systems that can be described by the independent or free particle model are determined by their single-particle energy spectra depending on the type of the fermion confinement and the size of the system. For such systems the heat capacity $C$ can be calculated as an integral involving the temperature variation of the occupation numbers $n(\varepsilon)$, i.e. $\partial n(\varepsilon)/\partial T$ (the temperature $T$ is measured in energy units), and the exact level density

\begin{equation}\label{rhoex}
\rho_{ex}(\varepsilon)=\sum_{t}\delta(\varepsilon-\varepsilon_t)d_t ,
\end{equation}
$\varepsilon_t>0$, $d_t\geq 1$ are respectively the energy and degeneration of each single particle level $t$:

\begin{equation}\label{C}
C/k_B=\int_0^{\infty}\rho_{ex}(\varepsilon)\varepsilon
\frac{\partial n(\varepsilon)}{\partial T}d\varepsilon; \;\;\;\;\;
\end{equation}
We shall consider $C$ for a fixed particle number $N$ that imposes on function $n(\varepsilon)$ the condition

\begin{equation}\label{eqN}
N=\int_0^{\infty}\rho_{ex}(\varepsilon)n(\varepsilon)d\varepsilon.
\end{equation}
Eq.~(\ref{eqN}) is fulfilled automatically if a fermion ensemble is treated  as the canonical one ($CE$) and $n(\varepsilon)$ is the canonical occupation number.  If $n(\varepsilon)$ is replaced by the Fermi-Dirac function $f(\varepsilon)$

\begin{equation}\label{FermiDir}
f(\varepsilon)=\left [ 1 + e^{\beta(\varepsilon-\lambda)}\right ]^{-1};\;\;\;\; \beta=T^{-1}
\end{equation}
then Eq.~(\ref{eqN}) defines the temperature dependent chemical potential $\lambda$.

The temperature independence of $N$ ($\partial N/\partial\beta=0$), Eq.~(\ref{eqN})), in which also as in Eq.~(\ref{C}) function $n(\varepsilon)$ is replaced by $f(\varepsilon)$, results in the representation of $C$ in the form

\begin{eqnarray}\label{Cegce}
C/k_B=\int_0^{\infty}\rho_{ex}(\varepsilon)\varphi(\varepsilon) d\varepsilon ,\\
\varphi(\varepsilon)=\beta^2(\varepsilon-\lambda-\beta\frac{\partial\lambda}{\partial\beta} )^2 e^{\beta(\varepsilon-\lambda)}\left [ 1+ e^{\beta(\varepsilon-\lambda)}\right ]^{-2}\label{phi}
\end{eqnarray}

Eq.~(\ref{Cegce}) allows the interpretation of the heat capacity as an average of the level density and the role of an averager is played by $\varphi(\varepsilon)$-function.

To stress the difference in the description with the temperature dependent $\lambda$ and  a fixed $N$ from the  grand canonical ensemble description, in which the independent variable is $\lambda$, we shall apply the term the equivalent grand canonical ensemble ($EGCE$)~\cite{goldstein}.

The simplest approach to the canonical description is the so called  grand canonical ensemble ($GCE$) approximation. This approximation accepts $f(\varepsilon)$ as the occupation probability with temperature dependent $\lambda$ (defined by  Eq.~(\ref{eqN})) but the temperature derivative of $\lambda$ ($\partial\lambda/\partial\beta$) is omitted in calculations of $C$ in contrast with the $EGCE$ description. The $GCE$ approximation gives the correct results at those temperatures when $\beta\partial\lambda/\partial\beta$ is very small as compared with $\lambda$, that e.g. leads to the correct linear in $T$ law for variations of $C$ in electron gas at low enough temperatures. However this approximation gives wrong results at $T\longrightarrow 0$ for systems with degenerated but partially filled (at $T=0$) levels and at $T\longrightarrow\infty$ for all systems.  We do not apply this approximation in our calculations and show wrong results for $C$ given by this method only in some figures in subsequent sections.

The polynomial method developed in our work, Ref.~\cite{kuzmenko}, makes canonical ($CE$) calculations of $C$ not more complicated than calculations corresponding to the $EGCE$ description. Therefore in all figures given below we represent only the canonical results though in several cases the $EGCE$ heat capacities are given to demonstrate to what extent $C(CE)$ and $C(EGCE)$ differ.

Since the canonical occupation distribution $n(\varepsilon)$ is much closer to the step-wise distribution than the Fermi-Dirac one (see e.g. Refs.~\cite{denton,schon,landsberg}) calculations of $C$ with $n(\varepsilon)$ and $f(\varepsilon)$ at low temperatures lead to different but like results. As a rule $EGCE$ calculations overestimate values of $C$ at low $T$ though with increasing $T$ this difference disappears. However the simplicity of the basic relationship for $C$ in the $EGCE$ method gives the unconditional advantage to the $EGCE$ method over the canonical one, Eqs.~(\ref{Cegce}),~(\ref{phi}), in interpretation of results. This feature of the $EGCE$ approach will be exploited in subsequent sections.

\begin{figure}
\scalebox{0.7}{\includegraphics{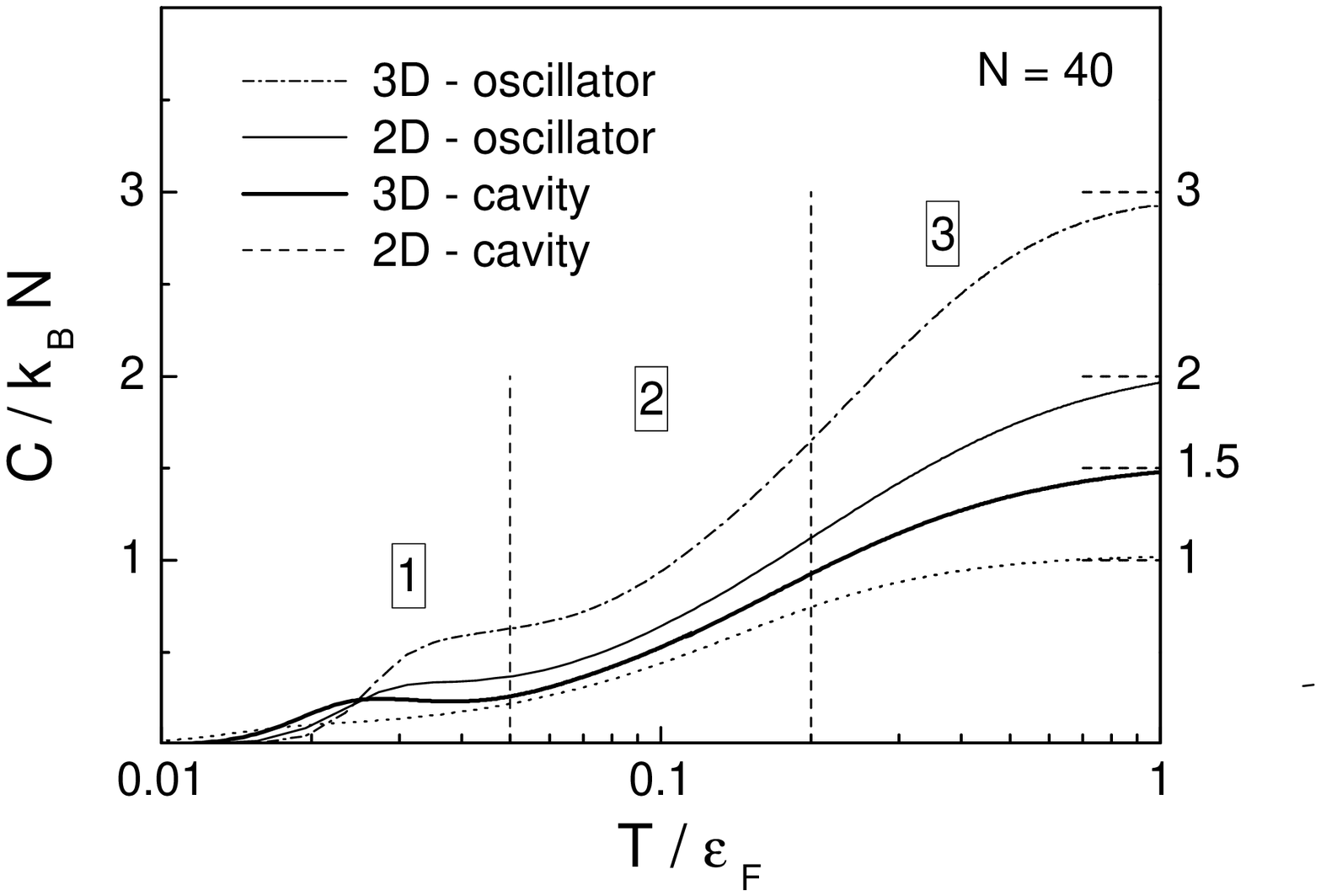}}
\caption{ The temperature dependence of the specific heat of $40$ fermions in $3D$-, and $2D$- isotropic oscillators ($\gamma=3;\;2$, respectively)  and spherical and circular rectangular wells ($\gamma=3/2;\;1$, respectively). Four stages of the temperature evolution of $C$ are distinguished. The first stage is the region of rapid temperature alterations of $C/k_BN$, it includes a local maximum in $C$. The second stage is a transition region from the rapid increase to the linear (or quasilinear) growth with $T$ that is the third stage. The fourth stage (not shown in the figure) is the region of the  saturation and transition to the classic limit of $C$ ($C/k_BN=\gamma$).}{\label{Allgamma40TEF}}
\end{figure}

The amount of levels contributed markedly to the heat capacity increases with $T$ giving rise to the growth of $C$. However the character of this growth is not monotonous and in the temperature evolution of $C$ from zero temperature to values of order of or more than  the Fermi energy ($\varepsilon_F$) four stages can be distinguished, Fig.~\ref{Allgamma40TEF}:                                                          
\begin{itemize}
\item The low temperature ($T<\delta_F$) exponential growth of $C$ up to a resonance temperature $T_0$ ( a local maximum in $C$ for some particle numbers).

\item A transition temperature region of inside which the rapid increase of $C$ is replaced by practically linear $T$-dependence of $C$.

\item The linear ( or quasilinear ) growth of $C$ with $T$ which is well known for the electronic heat capacity in bulk metals. This stage is continuously transformed into the next stage.

\item The final stage of the evolution of $C$ leading to the saturation: the heat capacity stops increasing with temperature and reaches the Boltzmann-Maxwell limit.

\end{itemize}
To illustrate the characteristic features of $C$ at each stage of the evolution we have performed canonical calculations of $C$ for diverse systems mentioned in the Introduction in a wide range of $N$ ($N<10^5$). In these calculations as an energy unit for measuring single-particle  energies and the temperature we employ the Fermi energy $\varepsilon_F$ of the considered system with a fixed $N$. Therefore our data are appropriate for any fermion system with the same $N$, shape, temperature and the type of confinement, however representation of these data on the absolute temperature scale requires the knowledge of the absolute values of $\varepsilon_F$ for each $N$. We believe that it is a detached problem and do not consider it here with the exception of a special model in Sec.5 that gives $\varepsilon_F(N)$ as a function of $N$ with respect to the corresponding bulk value, $\varepsilon_F(bulk)$.

Various types of confinement potentials we consider can be labeled by a quantity $\gamma$. For systems in cavities with reflecting walls (rectangular potentials) $\gamma=D/2$, $D$ being the spatial dimension, for all oscillator potentials $\gamma=D$. This quantity determines the specific heat in the classic limit, at $T/\varepsilon_F\geq 1$ ($C/k_BN=\gamma$). For $N\gg 1$ the value of $\gamma$ carries information concerning the type of confinement at linear variation of $C$ with $T$ ($\varepsilon_F N^{-1}\ll T\ll\varepsilon_F$ ) i.e. in the Sommerfeld heat capacity

\begin{equation}\label{Clin}
C_{lin}/k_BN=\gamma\frac{\pi^2}{3}\frac{T}{\varepsilon_F}.
\end{equation}
Sometimes, to stress peculiarities in $C$ of a system with a finite $N$ we adduce reduced values of the heat capacity $C(N)/C_{lin}$.

In some cases $\delta_F$, the mean level spacing for the Fermi gas with $N\gg 1$, is a more convenient unit to measure the temperature. :
\begin{equation}\label{deltaF}
\delta_F=\frac{d_F}{\rho_0(\varepsilon_F)}, \;\;\; \rho_0(\varepsilon_F)=\gamma N/\varepsilon_F
\end{equation}
$\delta_F$ includes the level degeneration $d_F$. For nonsymmetric systems $d_F=2$, for symmetric systems $d_F>2$. In particular, for spherical cavities the averaged degeneration $d_F$ is adopted to be equal to $\sqrt{N}$, Ref.~\cite{bohr},  $\rho_0(\varepsilon_F)$ being the smooth part of the level density at $\varepsilon_F$ and $N\gg 1$ (more complicated dependence of $\rho_0(\varepsilon_F)$ on $N$ for small $N$ discussed in Sec.5 is not essential here).

To simplify terminology we shall apply following names for systems under consideration. The term ``system in a cavity'' will imply a system of particles in a rectangular potential. If a cavity has a shape of parallelogram we shall distinguish two cases: a cube and a briquette. The latter is a parallelogram with different lengths of lateral ribs ($L_x\neq L_y\neq L_z$). The inscriptions in figures ``sphere'', ``cube'', ``briquette'' and so on imply that the values of $C$ are given for systems in cavities of corresponding shape, the inscription ``oscillator'' marks oscillator systems.

\section{Low temperature variations of the heat capacity}
\begin{figure}
\scalebox{0.7}{\includegraphics{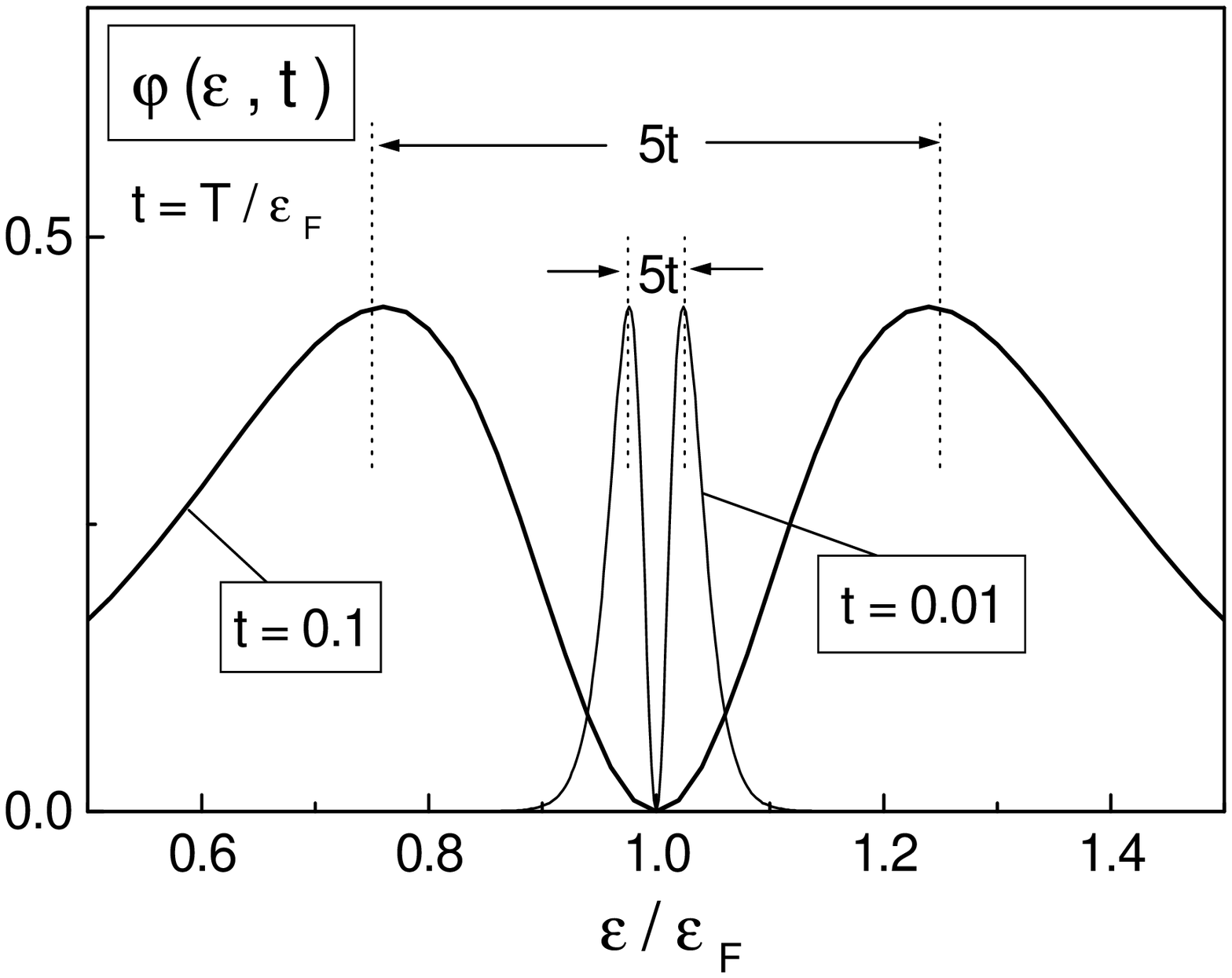}}
\caption{\label{twohump} The two humped function $\varphi(\varepsilon ,t)$, Eq.~(\ref{phi}), at $\beta\partial\lambda/\partial\beta=0$ and two temperatures: $t=0.01$ and $t=0.1$ ($t=T/\varepsilon_F$).}
\end{figure}

For macroscopic fermion systems the first two stages in the evolution of $C$ are absolutely inessential as the temperatures ($T<\delta_F$, $\delta_F\sim\varepsilon_F/N$) corresponding to these stages at $N\sim 10\:^{23}$ are so low that practically inaccessible. However for mesoscopic systems the first stage is of the most interest since the variations of $C$ v.s. $T$ can give information concerning single particle levels near $\varepsilon_F$ at temperatures accessible to measuring. (The absolute values of the temperature corresponding to the term ``low temperatures'' are determined by $N$ and the Fermi temperature which is material dependent: for metals $\varepsilon_F\sim 10^4\div10^5$~$K$  while for heterostructures $\varepsilon_F\sim 10^2$~$K$ and in trapped Fermi gases $\varepsilon_F\sim 1\mu K$.)  Such possibility is opened owing to the properties of the function $\varphi(\varepsilon)$, Eq.~(\ref{phi}). These properties are determined by two factors constituting $\varphi(\varepsilon)$; the first one is the bell-wise function

\begin{equation}\label{dfde}
e^{\beta(\varepsilon-\lambda)}\left [ 1+ e^{\beta(\varepsilon-\lambda)}\right ]^{-2}=f(\varepsilon)[1-f(\varepsilon)]=T\frac{\partial f(\varepsilon)}{\partial\varepsilon},
\end{equation}
with a maximum at $\varepsilon=\lambda$, the second factor is a parabola $\beta^{\,2}(\varepsilon-\lambda-\beta\frac{\partial\lambda}{\partial\beta} )^2$ with a zero minimum at $\varepsilon=\lambda+\beta\frac{\partial\lambda}{\partial\beta}$. Thereby $\varphi(\varepsilon)$ is a two humped function with the distance between two maxima $\approx 5T$, Fig.~\ref{twohump}. Such form of $\varphi(\varepsilon)$ confirms Grimvall's conclusion, Ref.~\cite{grimvall}, that the heat capacity probes the level density in an interval $\sim 10T$.

This distance weakly depend on $\beta\frac{\partial\lambda}{\partial\beta}$. If $\beta\frac{\partial\lambda}{\partial\beta}>0$ the left maximum is higher than the right one and vice versa at $\beta\frac{\partial\lambda}{\partial\beta}<0$. At high temperatures the left maximum does not take part in forming $C$ as $\lambda\ll 0$
and $\beta\lambda+\beta^2\frac{\partial\lambda}{\partial\beta}$ is positive and approximately equal to $\gamma$ (see Sec.6).

In order to understand how $\varphi(\varepsilon)$ forms values of $C$ at low temperatures it is useful to find the position of $\lambda$ with respect to the Fermi level ($F$). For this purpose we consider a model including only two levels: $F$ and $F+1$ supposing that $\varepsilon_F - \varepsilon_{F-1}$ and $\varepsilon_{F+2} - \varepsilon_{F+1}$ $>\varepsilon_{F+1} - \varepsilon_F$ (the notations $F\pm k$ correspond to the $k$ level above $(+)$ and below $(-)$ $F$ respectively).

The position of $\lambda$ strongly depends on $n_F$, the occupation number of $F$ at $T=0$:
\begin{equation}\label{lamop}
1\leq n_F\leq d_F-1;\;\;         \lambda\simeq\varepsilon_F - T\ln\left(\frac{d_F}{n_F}-1\right); \;\;\;
\lambda+\beta\frac{\partial\lambda}{\partial\beta}=\varepsilon_F;
\end{equation}

\begin{equation}\label{lamcl}
n_F=d_F;\;\;\;      \lambda\simeq\frac{1}{2}\left [(\varepsilon_F + \varepsilon_{F+1})+T\ln\frac{d_{F+1}}{d_F}\right]; \;\;\;
\lambda+\beta\frac{\partial\lambda}{\partial\beta}=(\varepsilon_F+\varepsilon_{F+1})/2.
\end{equation}
Corrections to Eqs.~(\ref{lamop},\ref{lamcl}) are proportional to $exp\left( -\beta\mid\varepsilon_{F\pm 1}-\varepsilon_F \mid\right )$ and omitted here.

Eqs.~(\ref{lamop},\ref{lamcl}) indicate that transition from the open shell to the closed one is accompanied by a drastic change of $\beta\partial\lambda/\partial\beta$ determining the minimum . Whereas at $n_F<d_F$ ($d_F\geq 2$) this quantity coincides with $\varepsilon_F$, i.e. the Fermi level does not contribute to $C$, at $n_F=d_F$ the value $\lambda+\beta\partial\lambda/\partial\beta$ is shifted to $(\varepsilon_F + \varepsilon_{F+1})/2$, i.e. $F$-level takes part in forming $C$ (~in both cases  $\beta\partial\lambda/\partial\beta$ does not practically depend on $T$).

It implies that this model ($EGCE$) predicts an attenuating factor, $exp[-0.5\beta(\varepsilon_{F+1}- \varepsilon_{F})]$, in the quotient of $C(n_F)$ and $C(d_F)$:

\begin{eqnarray}\label{Cnfdf}
C(n_F)/C(d_F)\simeq e^{-0.5x}\sqrt{\frac{d_{F+1}}{d_F}}\frac{2n_F}{d_F-n_F};\;\;\; n_F\leq d_F-1;\\
C(d_F)=\frac{1}{2} e^{-0.5x}x^2\sqrt{d_{F+1}d_F};\label{Cdf}\\
x=\beta(\varepsilon_{F+1}-\varepsilon_F). \nonumber
\end{eqnarray}

The canonical approach to the same model does not give such temperature attenuation but confirms the strong dependence on $n_F$

\begin{equation}\label{Cnf}
C(n_F;0\leq n_F\leq d_F)\simeq\frac{n_Fd_{F+1}}{d_F-n_F+1} e^{-x}x^2,
\end{equation}
i.e. $C(n_F=d_F)$ prevails over $C(n_F<d_f)$. Thus, C as a function of $N$ has to oscillate with $N$ and reveals maxima at $N$ corresponding to closed shells.

Heating extends $\varphi(\varepsilon)$ due to increasing the exponential factors in Eqs.~(\ref{Cdf},\ref{Cnf}) and approaching other levels ( in the first place $F-1$ and $F+2$) to the maxima of $\varphi(\varepsilon)$ that increases their contribution to $C$. These two factors stimulate growth of $C$.

A rather unexpected but quite natural consequence of the two humped character of $\varphi(\varepsilon)$ is the appearance of a resonance against a background of the monotonous growth of $C$ ( it is more evident in systems with high degenerations of levels). The resonance is conditioned by the coincidence of the points of the maxima in $\varphi(\varepsilon)$ with the energies of $F$ and $F+1$ levels. Since $5T$ is the distance between the maxima of $\varphi(\varepsilon)$ the resonance temperature is determined by the difference $\varepsilon_{F+1}-\varepsilon_F$:

\begin{equation}\label{T0}
T_0\simeq\frac{1}{5}(\varepsilon_{F+1}-\varepsilon_F).
\end{equation}
Such resonance amplification of $C$ in the low temperature region was predicted as far back as the work of Fr\"{o}lich ~\cite{frolich}. This local maximum can be found in the canonical calculations for $1D$- oscillator with $N\gg 1$, Ref.~\cite{denton}, and for alkali clusters, Ref.~\cite{brack}. However, the resonance nature of this phenomena is revealed here for the first time.

\begin{figure}
\scalebox{0.5}{\includegraphics{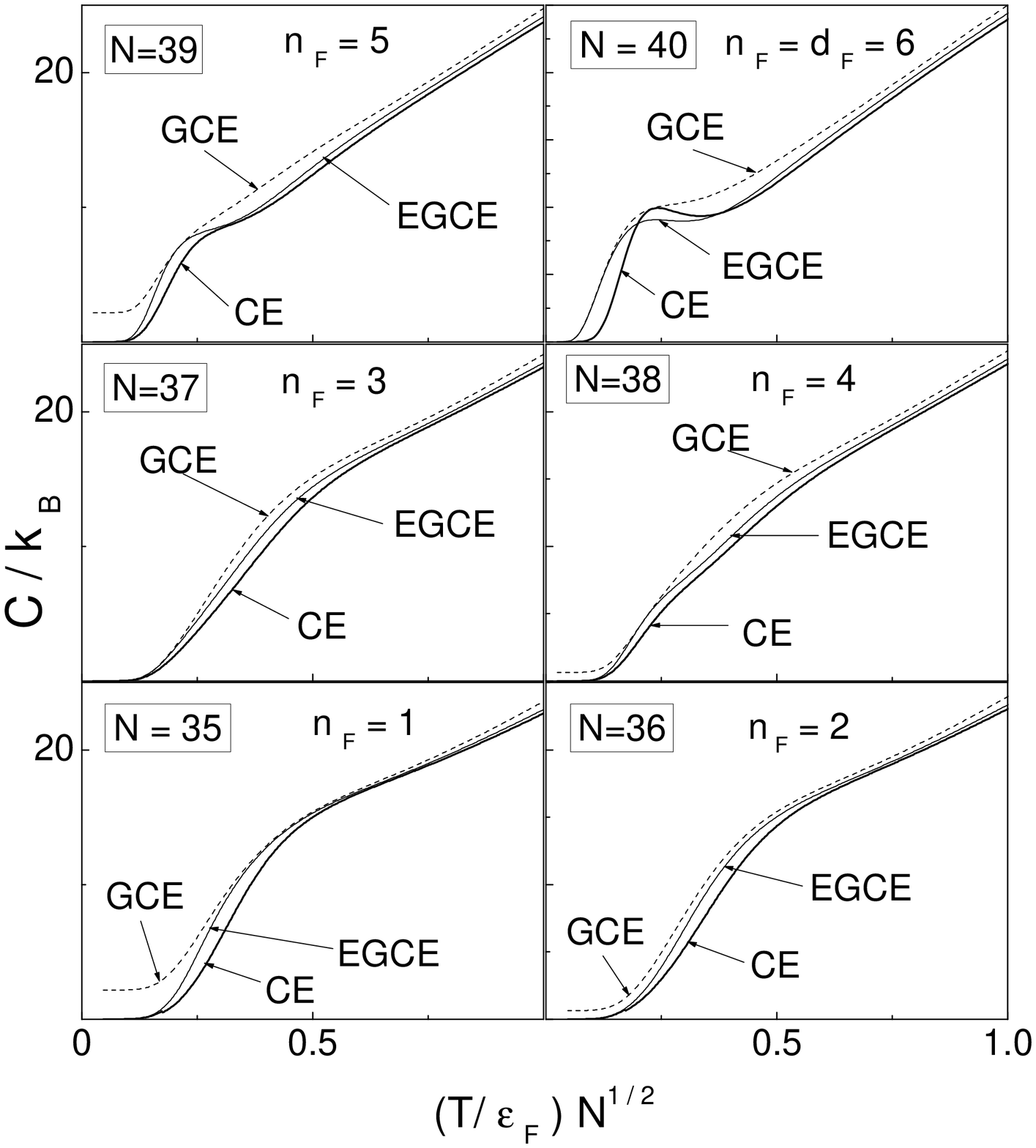}}
\caption{ Impact of the Fermi level filling  on the heat capacity for shell $N=40$ in the spherical cavity, $n_F$ is the particle number at $T=0$ on the Fermi level with degeneration $d_F=6$. Results of the canonical  ($CE$), equivalent grand canonical ($EGCE$) and grand canonical ($GCE$, $\lambda$ is temperature dependent, $\beta\partial\lambda/\partial\beta$ is omitted) methods are displayed. It is obvious that the $GCE$ method does not give correct values of $C$ at $T\longrightarrow 0$. In both other methods  $C\longrightarrow 0$ at $T\longrightarrow 0$.}{\label{C3540}}
\end{figure}
\begin{figure}
\scalebox{0.5}{\includegraphics{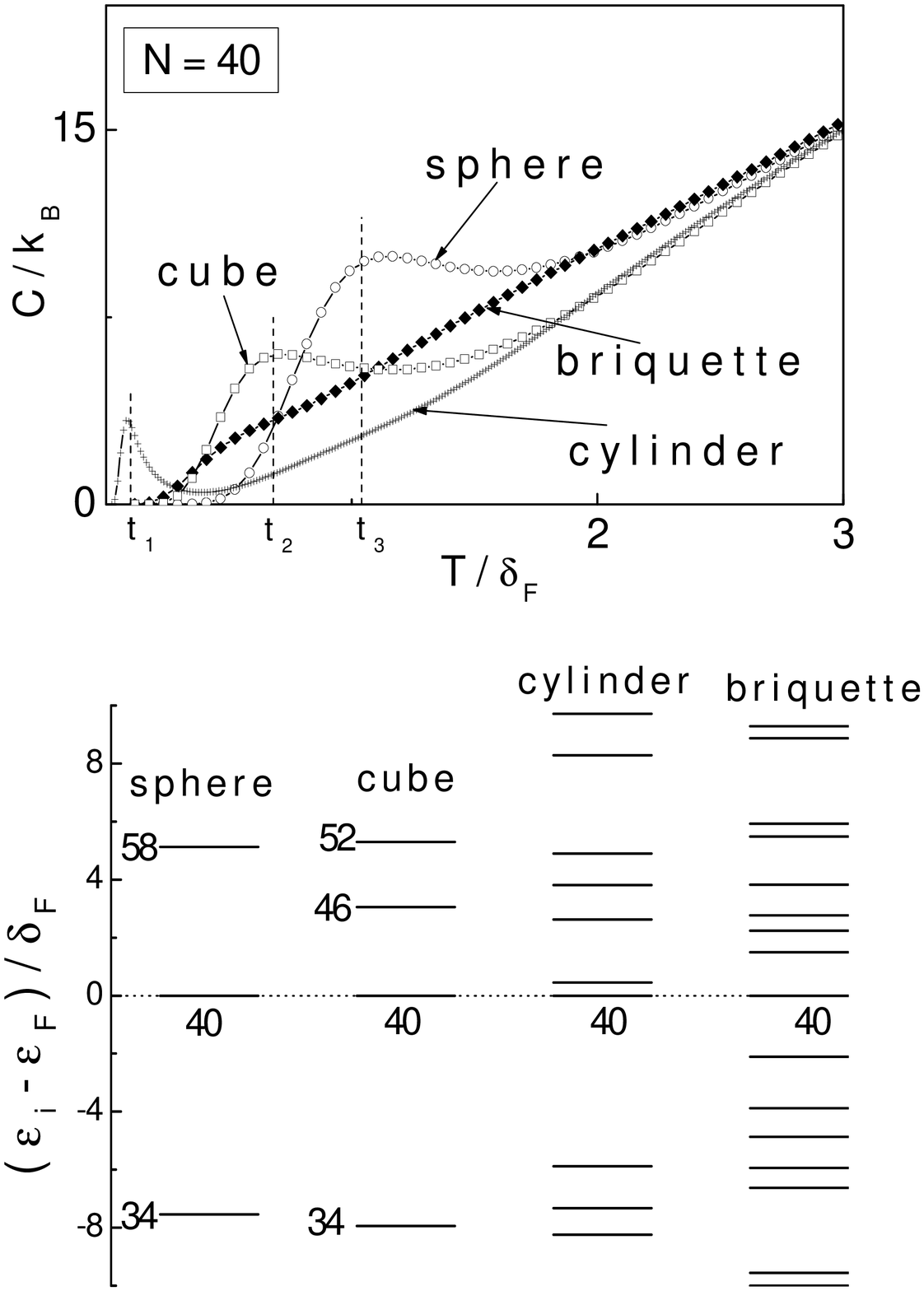}}
\caption{\label{LayCNa40cylsphbr}  Top panel: The heat capacities of  $N=40$ systems in $3D$-cavities of different shape (briquette, cube, cylinder and sphere ). The dashed vertical lines mark values of  $t=(\varepsilon_{F+1} - \varepsilon_F)/5\delta_F$ ( see Eq.~(\ref{T0})) which practically coincide with the positions of the corresponding maxima in $C$. In this figure $\delta_F=4\varepsilon_F/3N$. Bottom panel: Fragments of the single-particle level schemes of  sphere,  cube, cylinder( the diameter is equaled to the height, the level degeneracy in cylinder is equal to $2$ or $4$) and briquette   ($L_x:L_y:L_z=1:1.1:1.2$, the level degeneration is equal to $2$).}
\end{figure}
\begin{figure}
\scalebox{0.7}{\includegraphics{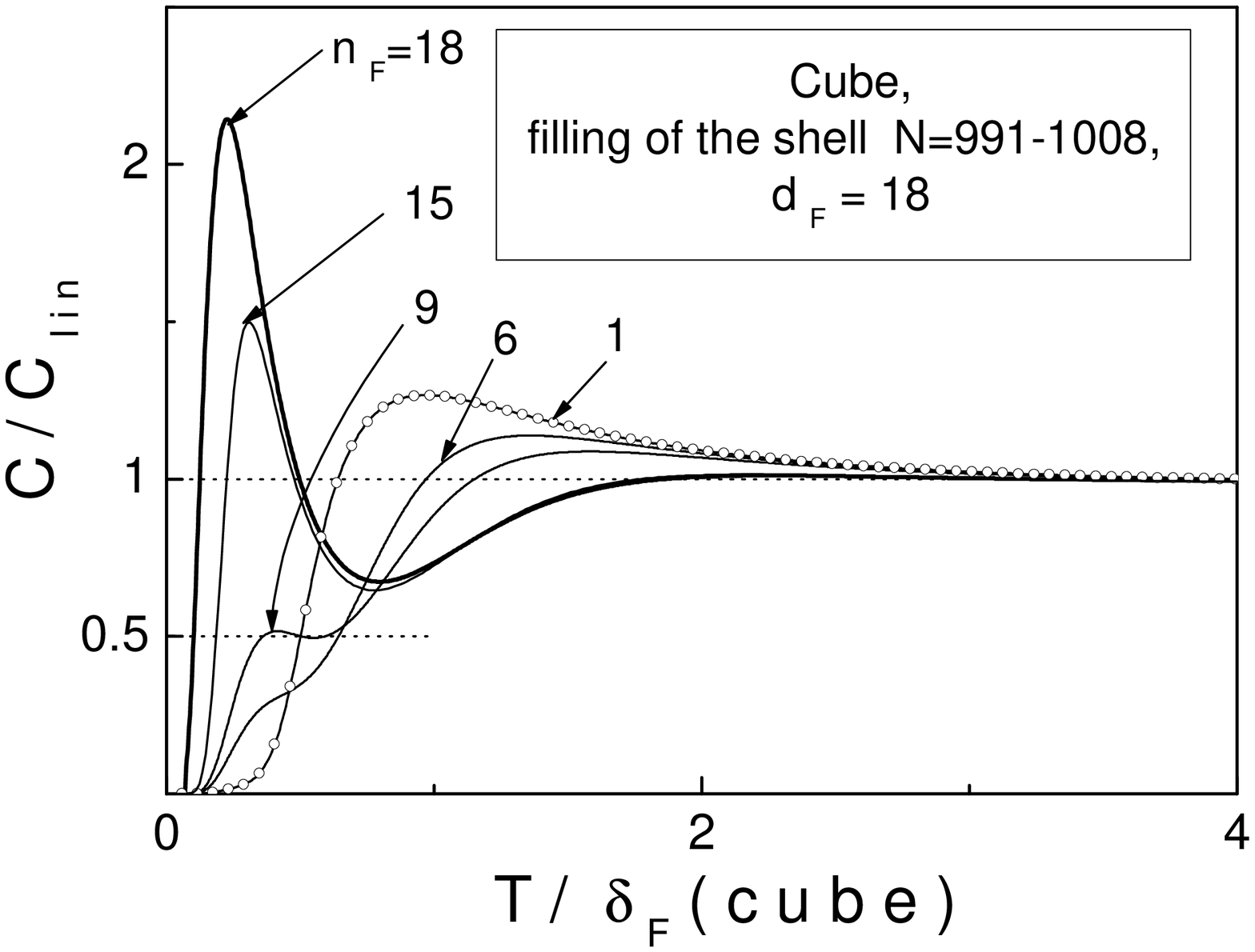}}
\caption{\label{CubedF18} Variations of the reduced   heat capacity ($C/C_{lin}$) vs the reduced temperature $\tau=T/\delta_F(cube)$ in  the degenerate shell of a cube, $\delta_F(cube)=d_F/\rho_0(\varepsilon_F)$, Eq.~(\ref{deltaF}). The averaged degeneration $d_F$ in a cube is adopted to be equal to $10$ ($d_F=10$).}
\end{figure}

Figs.~\ref{C3540},~\ref{LayCNa40cylsphbr},~\ref{CubedF18} indicate that such maximum occurs in various systems . In all cases in these figures the temperature of the maximum practically corresponds to Eq.~(\ref{T0}).

In symmetric systems (spherical and cubic cavities, spherical oscillators) these temperature resonances reach maximum amplitudes for closed shell (i.e. in even systems). However these resonances are high enough also at $n_F=d_F-1$. Development of such resonances with increasing $n_F$ is shown for spheres and cubes in  Figs.~\ref{C3540},~\ref{CubedF18}.

\begin{figure}
\scalebox{0.5}{\includegraphics{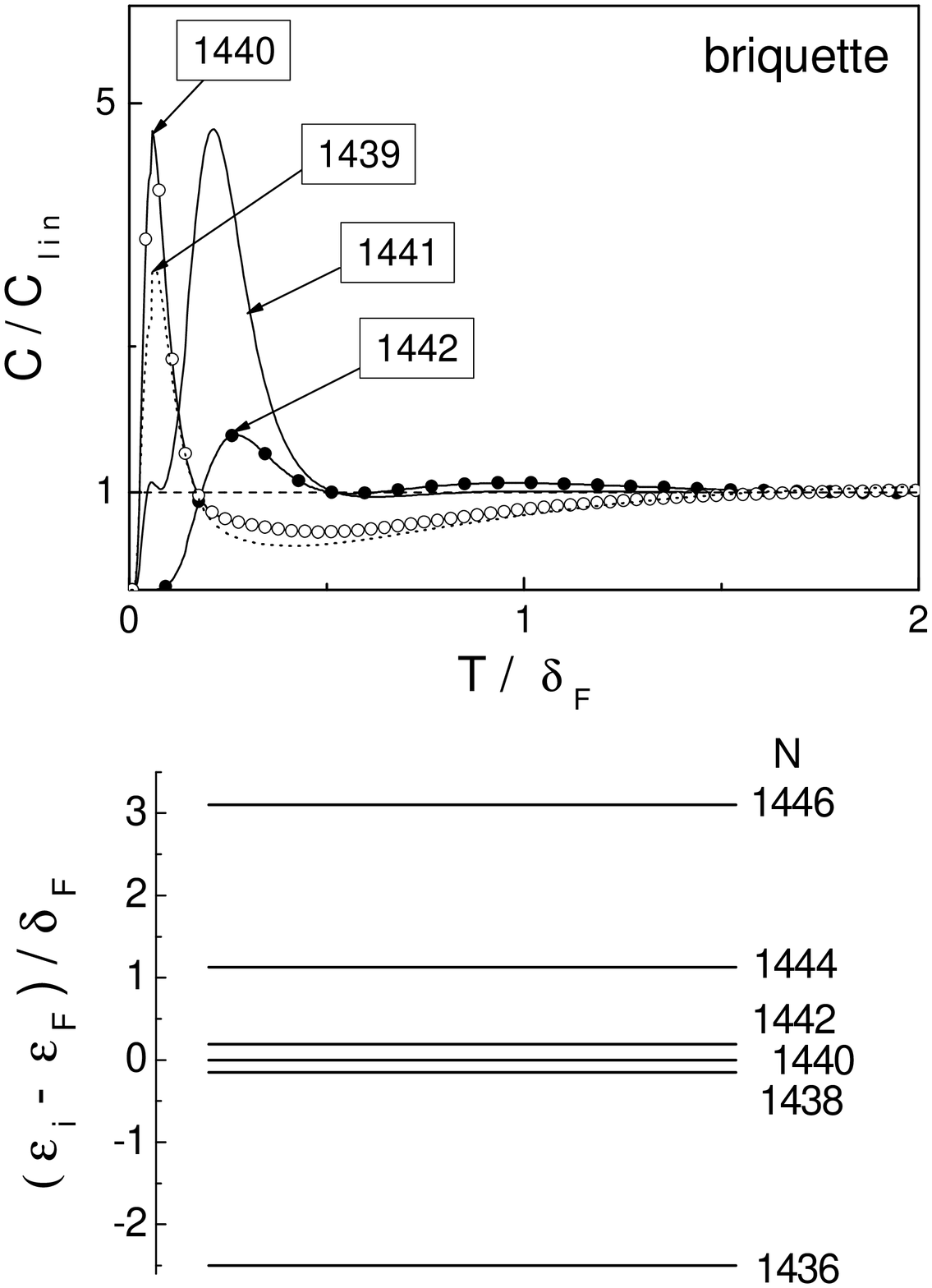}}
\caption{\label{Br14391442} Top panel: The reduced  heat capacity ($C/C_{lin}$) of $N$-even and $N$-odd systems vs $T/\delta_F$ for a briquette ($L_x:L_y:L_z=1:0.6e:\pi$), $\delta_F=4\varepsilon_F/3N$. Bottom panel: Fragment of the single-particle spectrum of the $N=1440$ briquette.}
\end{figure}

If in the vicinity of $F$ the spin degenerated ($d_F=2$) single-particle spectrum is uniform ($\delta_F$ is the level spacing) then, as shown by Denton et al in Ref.~\cite{denton}, the local maximum in $C$ at $T\sim 0.2\delta_F$ is explicitly expressed in even systems while in odd ones at $T\sim 0.4\delta_F$ there is only irregularity in increasing $C$. However for realistic spectra level concentration near $F$ can give maxima for both even and odd particle numbers. Fig.~\ref{Br14391442} indicates which variants can occur e.g. for briquettes with $N\sim 10^{\,3}$. Thus at a proper choice of $N$ studying the temperature dependence of $C$ can give the unique information concerning the difference $\varepsilon_{F+1}-\varepsilon_F$ in mesoscopic systems.
\begin{figure}
\scalebox{0.5}{\includegraphics{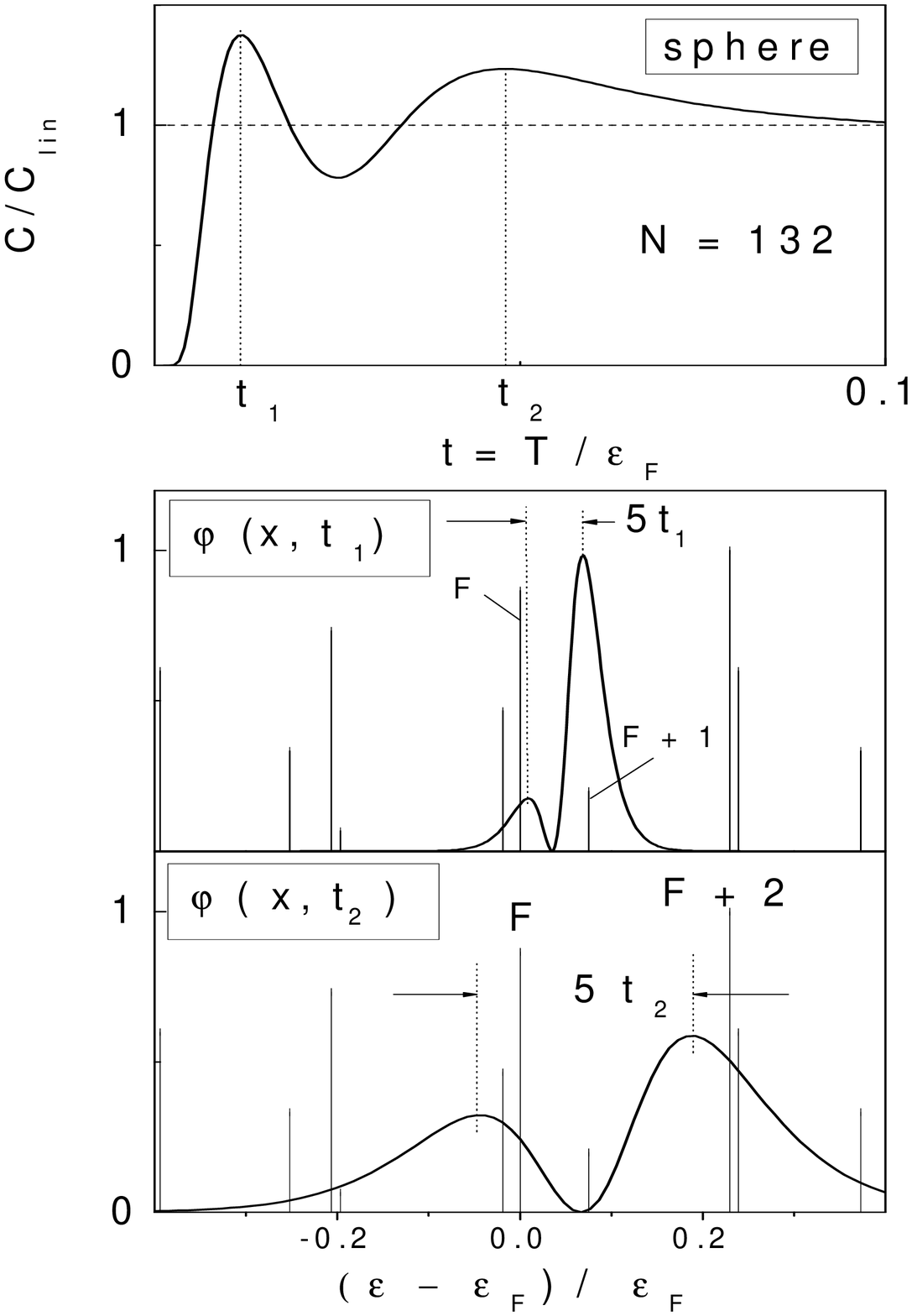}}
\caption{\label{2maxIntFunSph} Top panel: The reduced  heat capacity ($C/C_{lin}$) with two resonances for $132$ fermions in spherical cavity. $t_1$ and $t_2$ are the positions of maxima. Bottom panel: The two humped functions $\varphi(\varepsilon ,t)$,  Eq.~(\ref{phi}) at these temperatures. The first maximum is practically at $t_1=(\varepsilon_{F+1}-\varepsilon_F)/5\varepsilon_F$ but $t_2$ corresponds to the presence of two groups of levels in the vicinity of the maxima in $\varphi(\varepsilon ,t)$. Vertical lines mark the single-particle levels, their heights are proportional to the level degenerations.}
\end{figure}
\begin{figure}
\scalebox{0.5}{\includegraphics{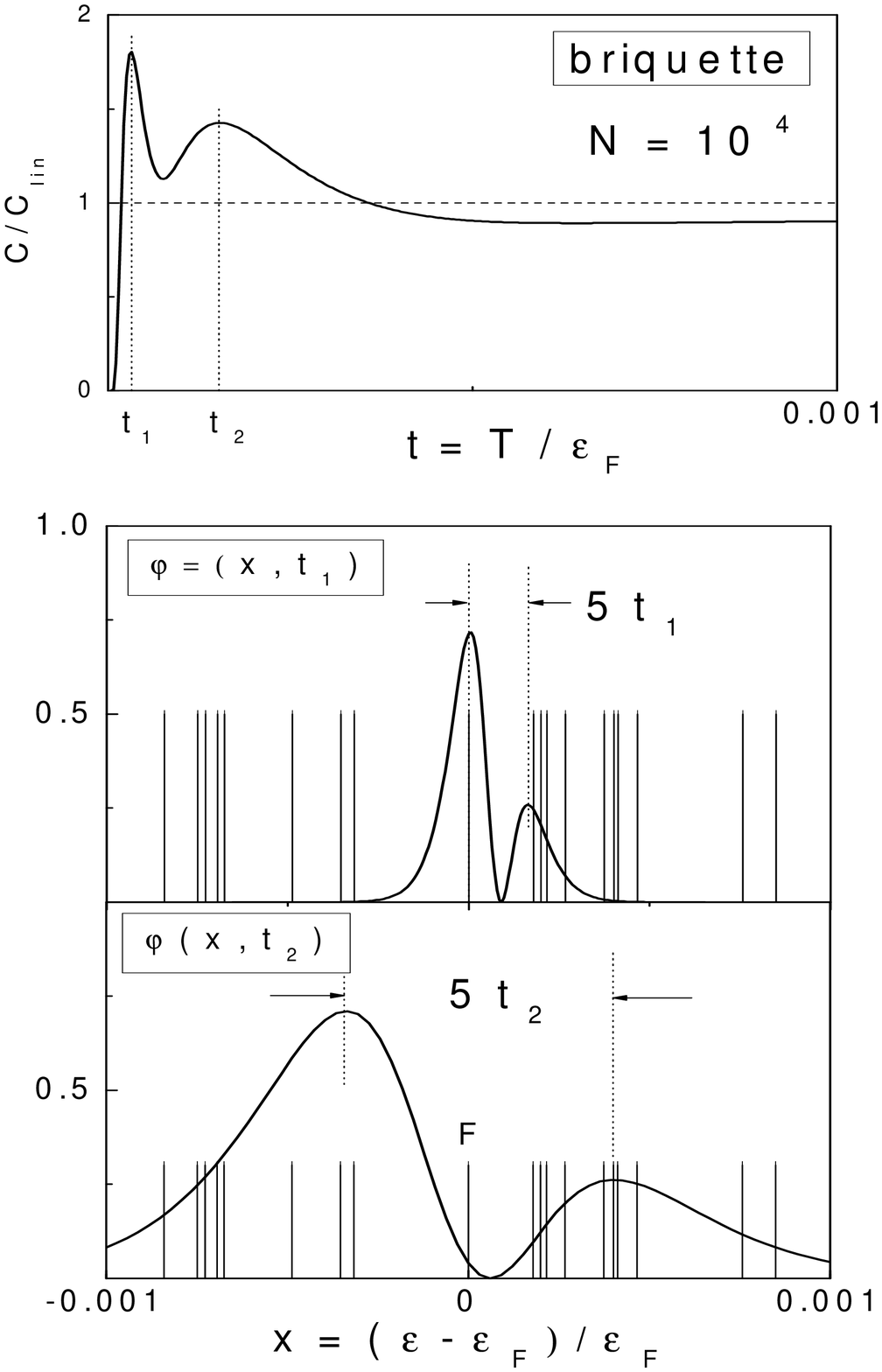}}
\caption{\label{2maxIntFunBr} The same as in Fig.~\ref{2maxIntFunSph} but for the $N=10^4$ briquette.}
\end{figure}
\begin{figure}
\scalebox{0.7}{\includegraphics{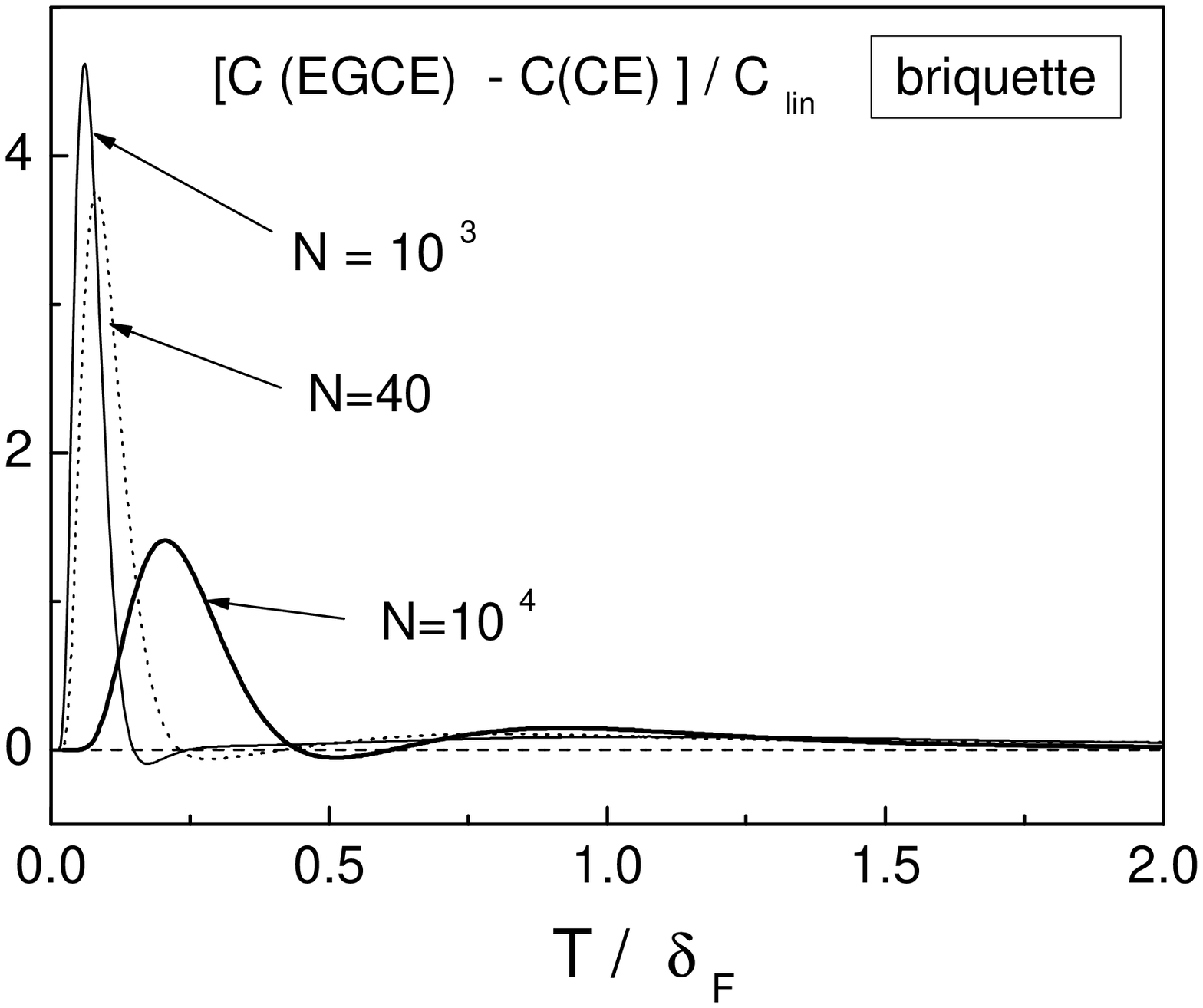}}
\caption{\label{CompareBr} Difference between  $C(CE)$ and  $C(EGCE)$ as a function of $T$ for briquettes ($\delta_F=4\varepsilon_F/3N$) with different $N$. The most divergence is observed in the region of the resonance temperatures.}
\end{figure}
\begin{figure}
\scalebox{0.5}{\includegraphics{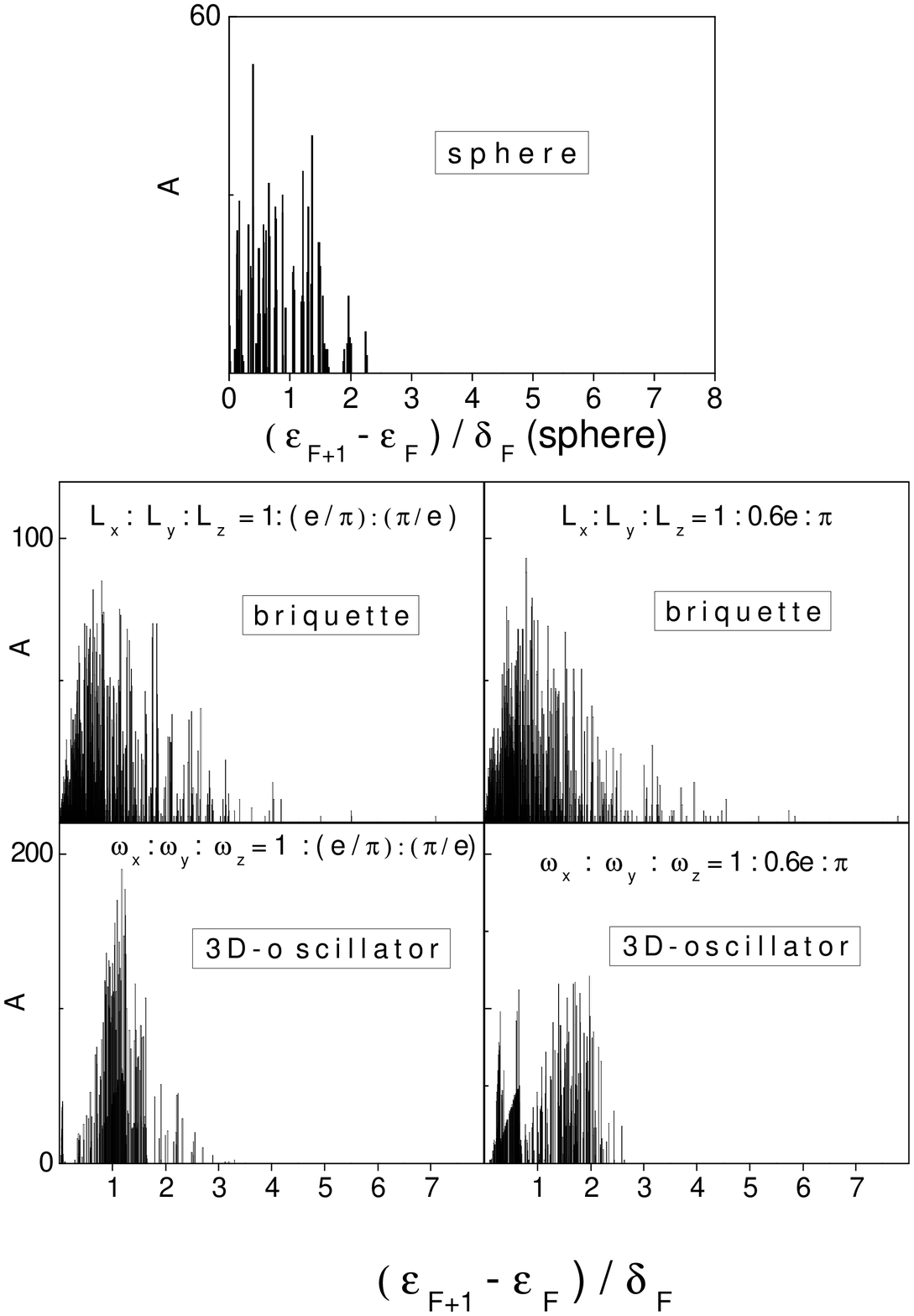}}
\caption{\label{LayDistrib} Distributions ($A$) of values $\Delta\varepsilon_{F+1,F}/\delta_F=(\varepsilon_{F+1} - \varepsilon_F)/\delta_F$ in different systems under consideration with $20 < N < 10^4$. $\delta_F$ is given by Eqs.~(\ref{deltaF}). The histograms for $A$ are built with the interval equal to $0.02$. We use transcendental numbers for the ratio of ribs in briquettes and oscillator frequencies to avoid accidental level degenerations.}
\end{figure}

In systems  with nonuniform spectra this difference can turn out to be small as compared with other level differences near $F$. In this case increasing $T$ can give one more maximum in $C$. Examples of such cases are in Figs.~\ref{2maxIntFunSph},~\ref{2maxIntFunBr} which indicate that the second resonance is formed by not one pair of levels, as it takes place for the first resonance ($5T_0=\Delta\varepsilon_{F+1,F}=\varepsilon_{F+1}-\varepsilon_F$) but in this case under the maxima of $\varphi(\varepsilon)$ there are at least two groups of levels (bunches).

It should be mentioned that in the region of the first temperature resonance the divergence between $C(CE)$ and $C(EGCE)$ becomes maximal and then with increasing $T$ this difference decreases and practically disappears at $T>2\delta_F$, Fig.~\ref{CompareBr}.

To assess which differences  $\Delta\varepsilon_{F+1,F}$ can occur in fermion mesoscopic systems and accordingly which temperature gives rise  to resonances in $C$ we have analyzed these differences in diverse systems with $N$ ranged from $20$ to $10^4$. These data are displayed in Fig.~\ref{LayDistrib} in units $\delta_F$, Eq.~(\ref{deltaF}).

For the spherical oscillator the resonance temperature is strictly fixed ($T_0\simeq\omega/2$) but in spherical cavities there is a rather wide interval of possible values of $T_0$. This interval would be much wider if one takes the same $\delta_F$ (the denominator of $x$ ) as in nonsymmetric systems i.e. without allowing for the averaged degeneration in spherical cavities. Possible additional degenerations have to be taken into account also at appearance of new types of symmetries arising when the ratios of the lateral lengths in briquettes or the frequencies in oscillator systems become equal to the ratios of small integers.

For nonsymmetric systems in cavities the distributions of $x=(\varepsilon_{F+1}-\varepsilon_F)/\delta_F$ weakly vary with deformation of the system   (excepting cases that lead to a new symmetry). However they evidently differ for $3D$-cavities and $3D$-oscillators. Nevertheless these distributions in both cases are rather wide. Therefore maxima in $C$ can occur at temperatures extending from very small values $\ll\delta_F$ up to $(2\div 3)\delta_F$.

We would like to stress that the distributions in Fig.~\ref{LayDistrib} are not those considered by Kubo and his followers. Kubo et al took into account the distributions of all spacings in an $N$-particle system. Whereas in Fig.~\ref{LayDistrib} the distributions of the only kind of spacings, $(\varepsilon_{F+1}-\varepsilon_F)$ are presented but the particle numbers are in a very wide range $20<N<10^{\,4}$ at each type of confinement.

After temperature resonances which complete the first stage of the evolution of $C$ thermodynamic properties of a heated system are determined by several single-particle levels placed under maxima of $\varphi(\varepsilon)$, i.e. beginning with some temperatures the main role in forming $C$ belongs to the smooth level density.


\section{Transition to the regime of the smooth level density}

As mentioned above at $T$ higher than the first or second resonance temperature several single particle levels ($>2$) appear under and near the maxima of $\varphi(\varepsilon)$ that starts on averaging the level density $\rho(\varepsilon)$. This process is ended at the temperature $T_{sm}$, the temperature smoothing the level density oscillations, and accompanied by such interesting quantum size effect as oscillations of the heat capacity v.s. $N$. This phenomenon results in the variations of the specific heat $C/N$ v.s. $N$, i.e. it exhibits one more paradoxical property of mesoscopic fermion systems in comparison with macroscopic ones.

As well known the exact $\rho(\varepsilon)$, Eq.~(\ref{rhoex}), can be represented as a sum of a smooth function of energy $\rho_0(\varepsilon)$ and an oscillating shell correction $\delta\rho(\varepsilon)$

\begin{equation}\label{rhotot}
\rho(\varepsilon)=\rho_0(\varepsilon)+\delta\rho(\varepsilon)
\end{equation}
For simplest single-particle 3D-rectangular potentials values of $\rho_{\,0}(\varepsilon)$ were established many years ago and these results have been collected by Balian and Bloch ~\cite{balian}. For $3D$-oscillators $\rho_0(\varepsilon)$ is given by Bohr and Mottelson~\cite{bohr}. The component $\delta\rho(\varepsilon)$ is an infinite set of functions oscillating with $\varepsilon$ the period of which decrease as $1/n$ ($n=1,2,\ldots$). Each separate oscillation in $\delta\rho(\varepsilon)$ with a period $\tau_n$ will be named $n$-mode.

In the previous section to interpret values of $C$ we use the narrow distribution of $\varphi(\varepsilon)$~-function at low temperatures to choose some levels determining $C$. In this section because of dilation of  $\varphi(\varepsilon)$ its role is changed and here it serves as an smoothing function to damp $n$-modes. In fact, if between the maxima of  $\varphi(\varepsilon)$ there are $\sim 3$ periods $\tau_n$, this mode will be suppressed and it will not practically contribute to $C$. It implies that the temperature $T_n$ which removes $n$-mode and period $\tau_n$ are connected by the condition
\begin{equation}\label{tauT}
5T_n\sim 3\tau_n,
\end{equation}
i.e. $T_n$ is of order of $\tau_n$. If all $\tau_n$ are proportional to a highest period $\tau_0$, as it takes place in spherical oscillators
\begin{equation}\label{rhoosc}
\rho(\varepsilon)=\frac{1}{\omega^3}(\varepsilon^2-\frac{\omega^2}{4})\left [ 1+ 2\sum_{n=1}(-1)^n \cos\frac{2\pi n }{\omega}\varepsilon\right ],
\end{equation}
$T_{sm}$, smoothing all $n$-modes, is determined by this maximum period (for the spherical oscillator it is $\omega$)
\begin{equation}\label{Ttau0}
T_{sm}\sim\tau_0 .
\end{equation}
For asymmetric oscillators (all frequencies are different) the highest possible period amounts to the most frequency of $\omega_x$, $\omega_y$, $\omega_z$. However if a system is moderately deformed ($\omega_x\sim\omega_y\sim\omega_z\sim\omega=(\omega_x\omega_y\omega_z)^{1/3}$ the value of $\omega$ can serve as an estimate of such period. The oscillator frequency can be bound to physical parameters of the system by different ways. One of them consists in equating the oscillator Fermi energy with that for systems in a rectangular potential. In another way the empirical root-mean square radii (for $3D$-systems $R\sim N^{1/3}$, for $2D$-systems $R\sim N^{1/2}$) are supposed to be equal to those calculated with oscillator functions (~\cite{bohr}). In both cases $\omega\sim\varepsilon_F N^{-1/3}$ ($3D$-systems) or $\omega\sim\varepsilon_F N^{-1/2}$ ($2D$-systems). For closed
oscillator shells with the main quantum number $\aleph_F$ and spin degenerated levels the exact relation between $\aleph_F$ and $N$ is known: $3N=(\aleph_F+1)(\aleph_F+2)(\aleph_F+3)$ for $3D$-systems and $N=(\aleph_F+1)(\aleph_F+2)$ for $2D$-systems. Hence one obtains for $\aleph_F\gg 1$ that $\omega\approx\varepsilon_F(3N)^{-1/3}$ and $\varepsilon_F(N)^{-1/2}$ respectively.

In spherical cavities the maximum period $\tau_0$ corresponds to oscillations created by supershells which are well studied ~\cite{bohr,gutzwiller,balian2,nishioka}. Since the length of supershells $\sim\varepsilon_FN^{-1/3}$  period $\tau_0$ of density oscillations in mesoscopic spheres takes the same value. That is in agreement with the theoretical estimate of $\tau_0$

\begin{equation}\label{tau0sph}
\tau_0\approx\left(\varepsilon_F\frac{\hbar^2}{2mR^2}\right)^{1/2},
\end{equation}
$R$ being the radius of the sphere, $R\sim N^{1/3}$.

Period $\tau_0$ for parallelepipeds is determined by the minimal lateral length of $L_x$, $L_y$, $L_z$. For systems with                           $L_x\sim L_y\sim L_z\sim L=(L_xL_yL_z)^{1/3}$ the value of $\tau_0$ is of the same order as for spheres:
\begin{equation}\label{tau0brik}
\tau_0\approx\left(\varepsilon_F\frac{\hbar^2}{2mL^2}\right)^{1/2}\sim\varepsilon_F/N^{1/3}.
\end{equation}
As far as high frequency modes are suppressed by heating the character of the $N$-oscillations is altered.
\begin{figure}
\scalebox{0.5}{\includegraphics{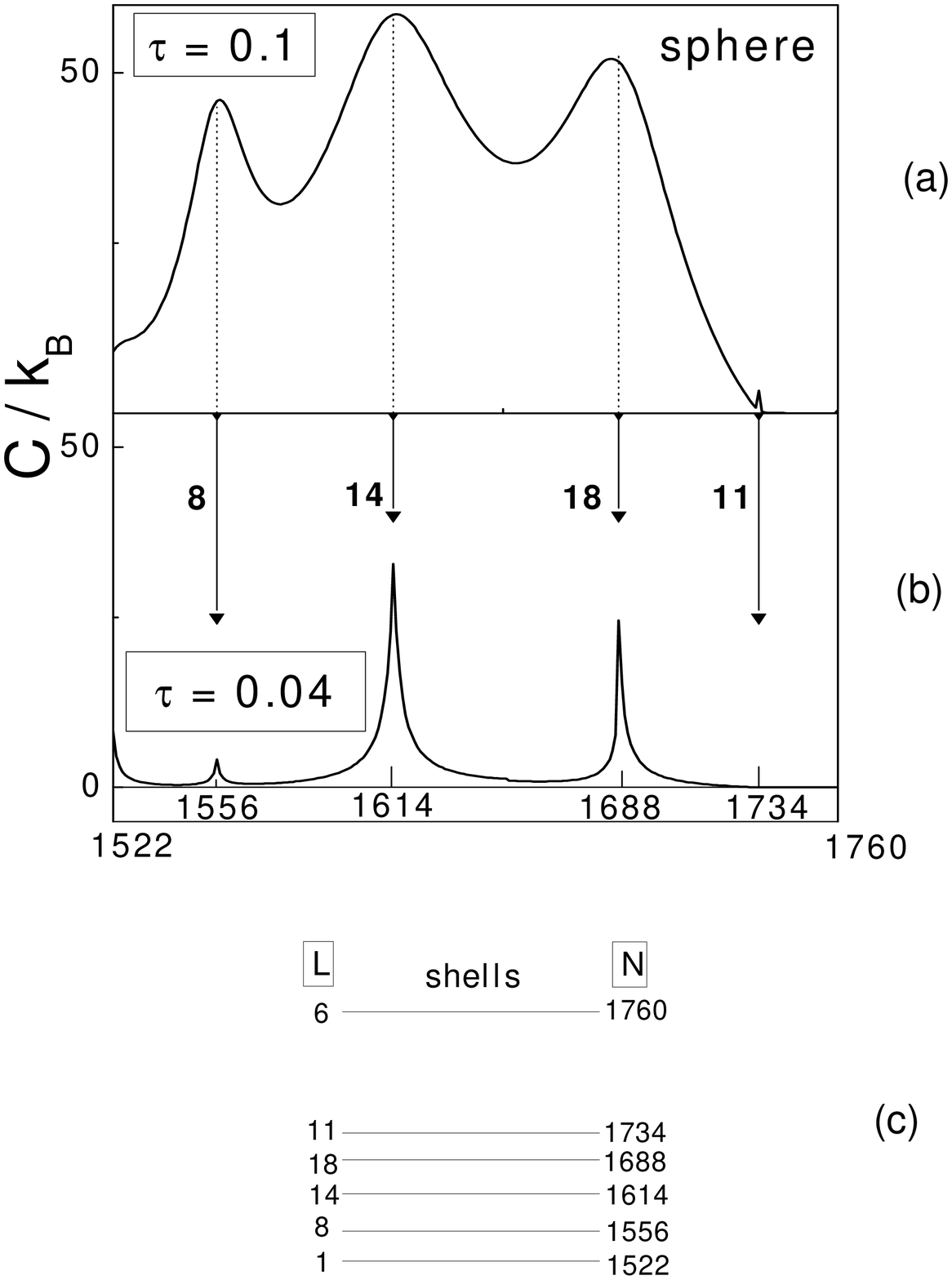}}
\caption{\label{Lay2Peaks} Top panels: Temperature development of $N$- oscillations of the heat capacity in spheres  vs $N$,  $\tau=(T/\varepsilon_F)N^{1/2}$. Bottom panel: Fragment of the level scheme of the spherical cavity.  $N$ is the particle number of a system with the closed Fermi shell.  $L$ is the orbital momentum of the Fermi shell.}
\end{figure}
\begin{figure}
\scalebox{0.7}{\includegraphics{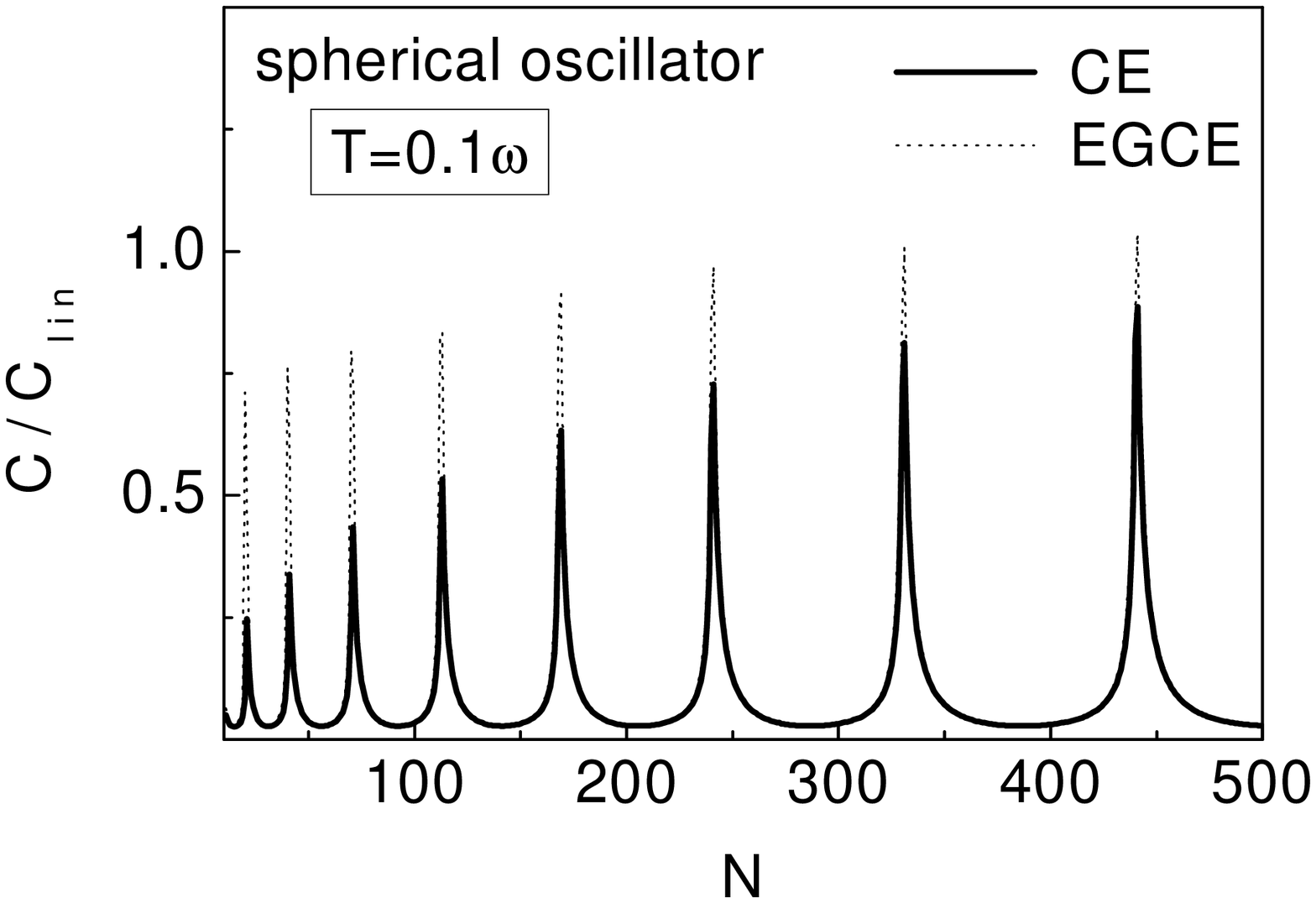}}
\caption{\label{C3DoscCompTeF01} $N$ - oscillations of the canonical heat capacity $C(CE)$ (bold line) and effective grand canonical one $C(EGCE)$ (dotted line)   at fixed $T/\omega=0.1$, $\omega$ is the spherical oscillator frequency.}
\end{figure}
\begin{figure}
\scalebox{0.5}{\includegraphics{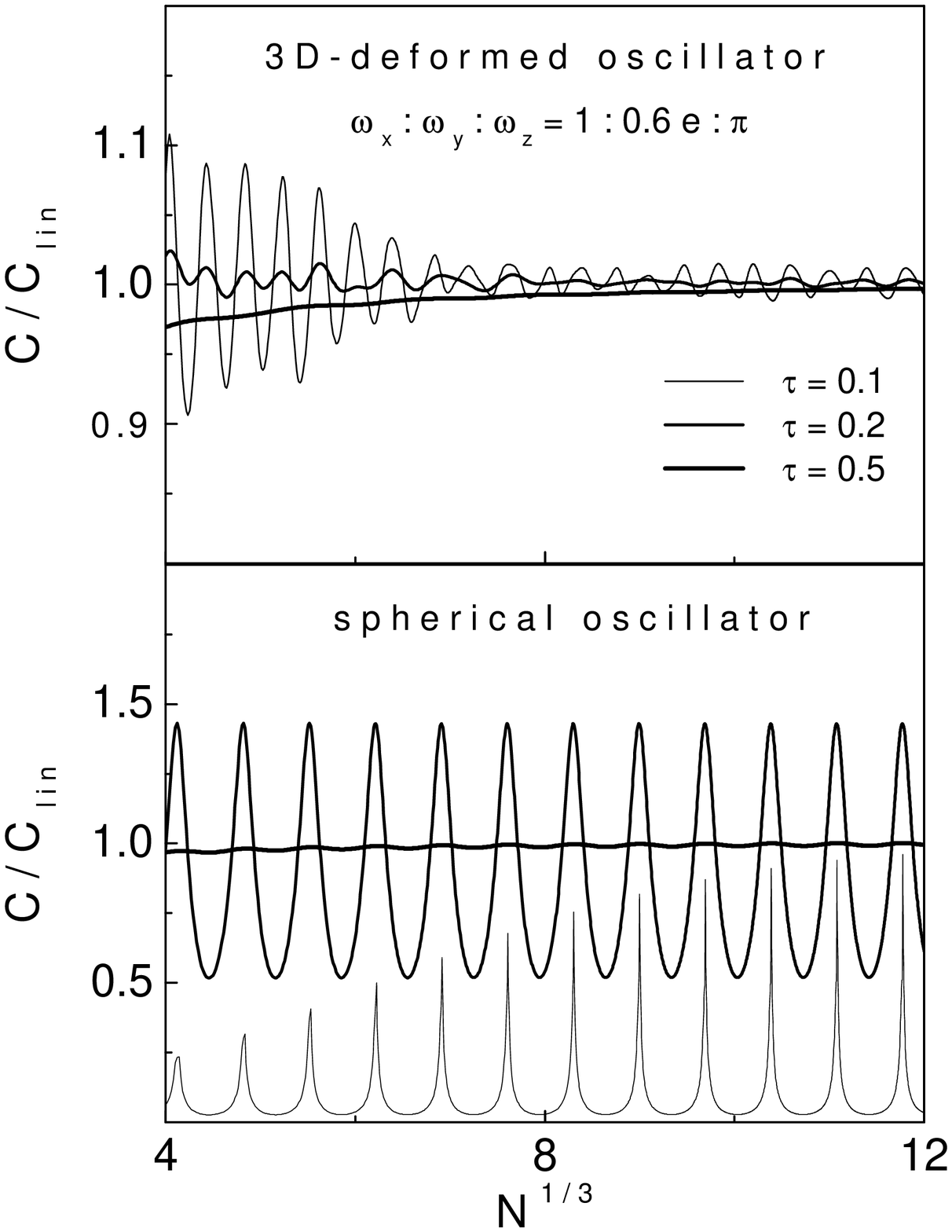}}
\caption{\label{CSphDefOsc} $N$ - oscillations of the reduced  ($C/C_{lin}$) heat capacity  at different temperatures $\tau=(T/\varepsilon_F)N^{1/3}$ for $3D$- deformed (top panel) and spherical (bottom panel) oscillators. Amplitudes of $N$-oscillations in deformed oscillators are considerably lesser than in spherical ones.}
\end{figure}
\begin{figure}
\scalebox{0.7}{\includegraphics{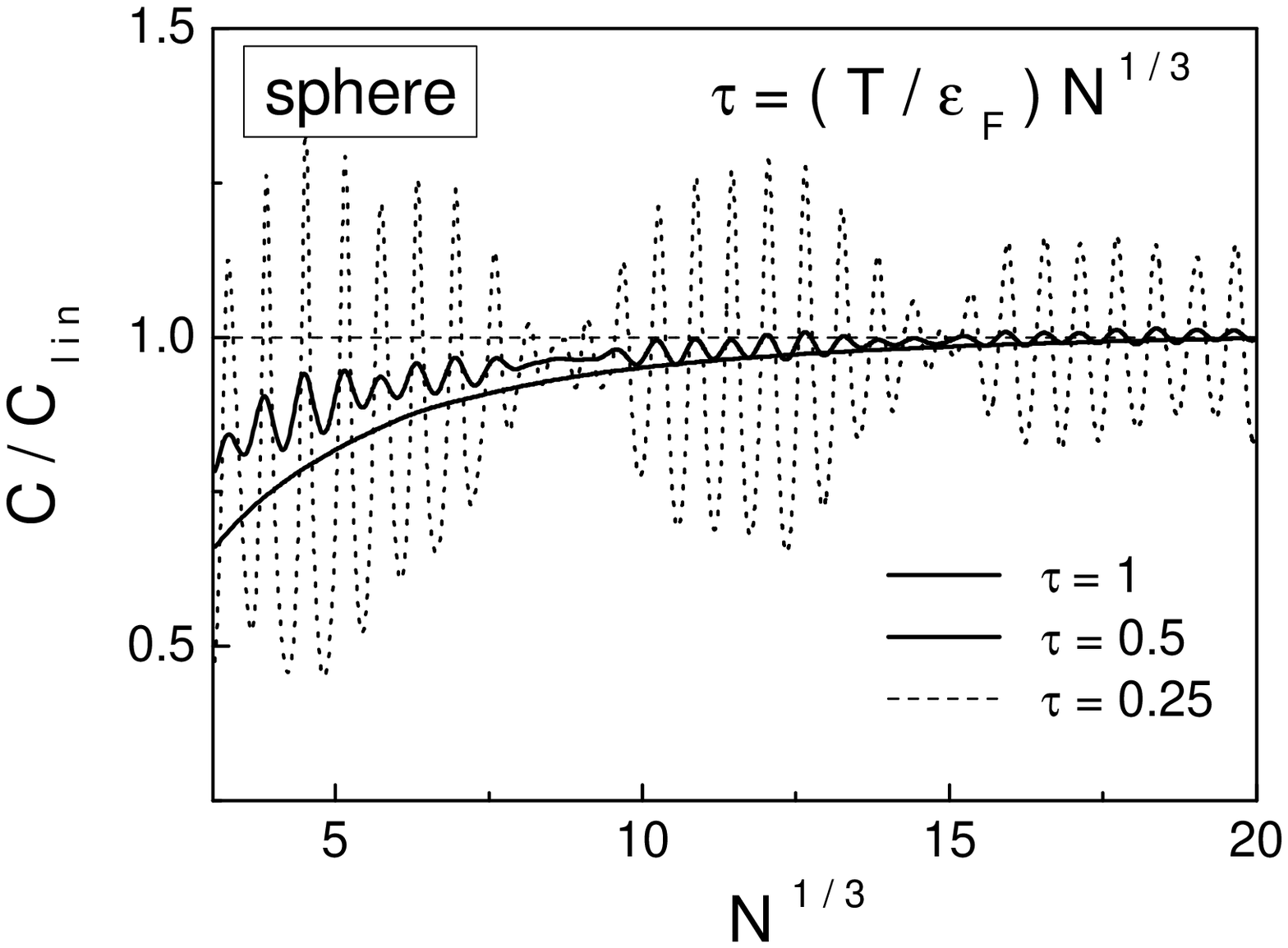}}
\caption{\label{CDsph025051} $N$- oscillations of the heat capacity ($C/C_{lin}$) in spheres at different reduced temperatures $\tau=(T/\varepsilon_F)N^{1/3}$.}
\end{figure}
\begin{figure}
\scalebox{0.5}{\includegraphics{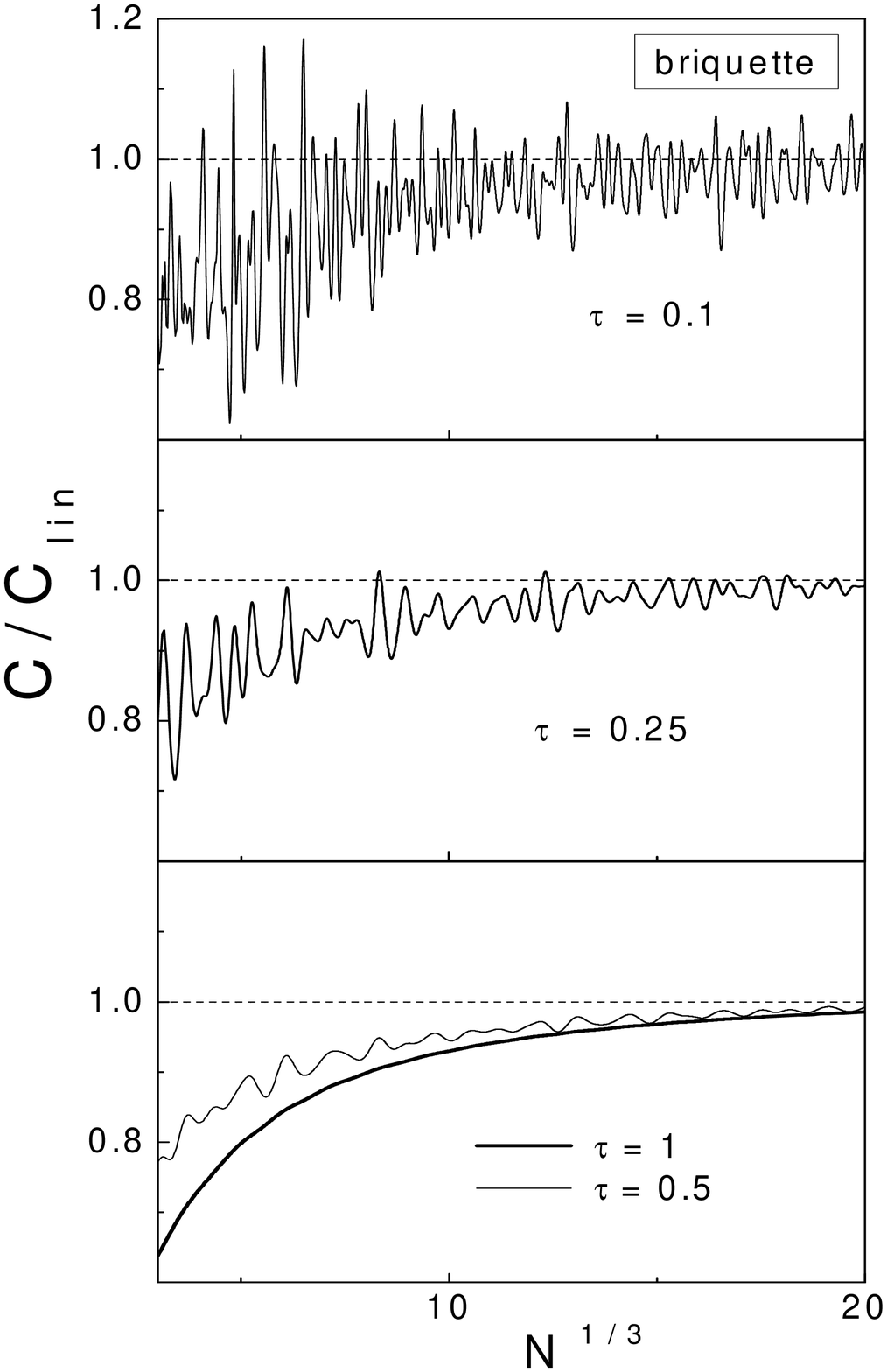}}
\caption{\label{CBrOsctau} $N$- oscillations of the heat capacity ($C/C_{lin}$) in briquettes at different reduced temperatures $\tau=(T/\varepsilon_F)N^{1/3}$.}
\end{figure}
\begin{figure}
\scalebox{0.5}{\includegraphics{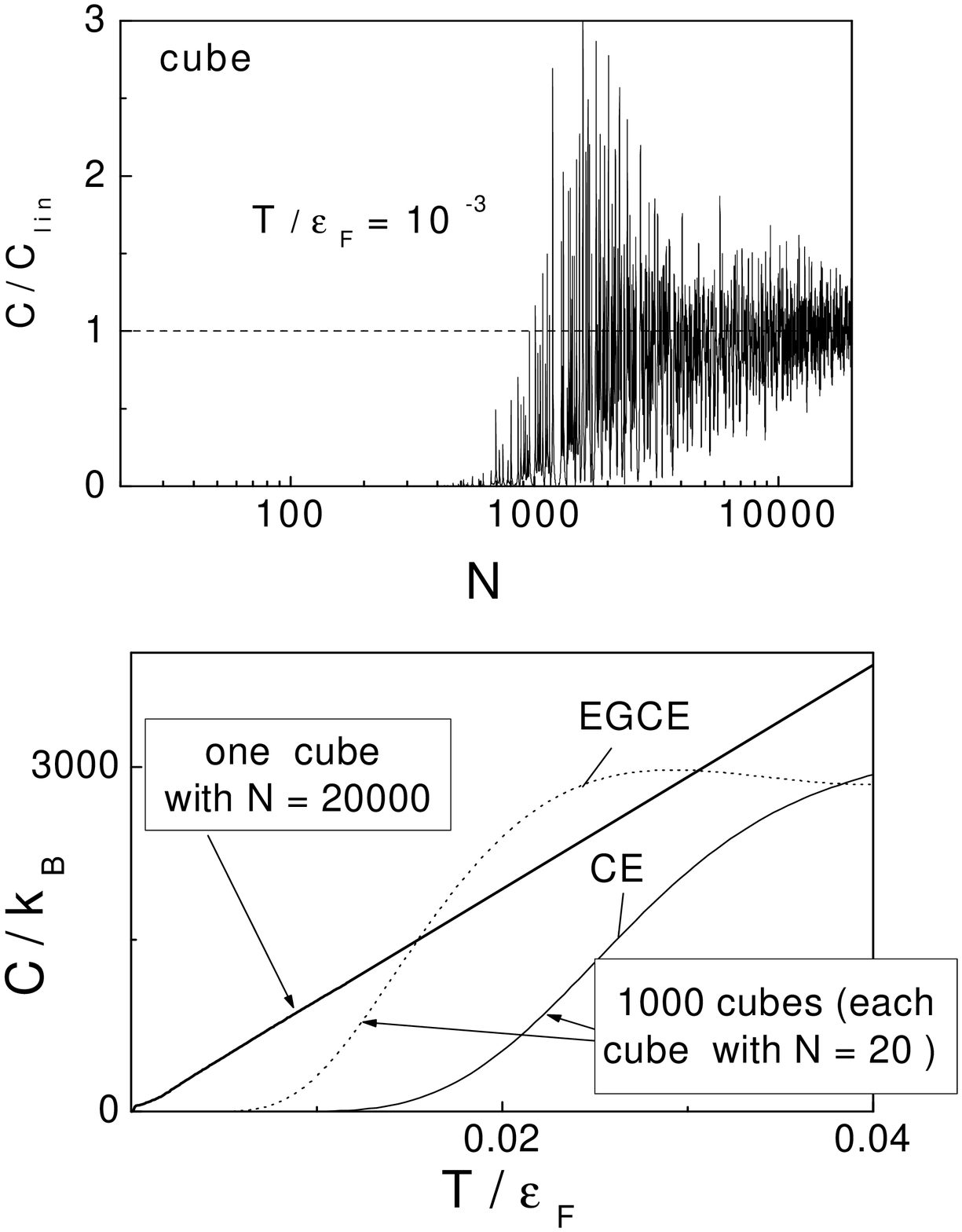}}
\caption{\label{Rubik} Top panel: $N$- oscillations of the heat capacity ($C/C_{lin}$) in cubes. Bottom panel: The comparison of the heat capacity of a cube with $N=20000$ with the total heat capacity of $1000$ small cubes (each with $N=20$).}
\end{figure}
At small $T$, as pointed out in Sec.3, $C$ vs $N$ at a fixed $T$ demonstrates maxima at each $N$ that corresponds to a filled level or shell. Amplitudes of these peaks are proportional to degenerations of adjacent levels. The origin and temperature development of several such peaks are displayed in Fig.~\ref{Lay2Peaks}. The physical nature of the difference in peaks consists in variations of the Fermi level occupation numbers in adjacent systems at $T\simeq 0$. This difference has to be attenuated with growing temperature which stimulates regularization of the peaks due to removing high frequency modes in $\delta\rho(\varepsilon)$. The gradual temperature refinement of the $N$-oscillations is shown in  Figs.~\ref{C3DoscCompTeF01},~\ref{CSphDefOsc},~\ref{CDsph025051},~\ref{CBrOsctau} for diverse systems (spherical and deformed oscillators, spherical and briquette cavities). These figures display such property of mesoscopic systems (in comparison with macroscopic ones) as the dependence of $C$ on the shape of the system ( even at a fixed type of confinement) and variations of the specific heat $C/N$ with $N$.

As an illustration to the latter statement in Fig.~\ref{Rubik} we compare the heat capacities of two systems containing $N=2\cdot 10^{\,4}$ fermions, e.g. electrons in a metal grain. The first is a cube with the rib $\sim(2\cdot 10^{\,4})^{1/3}r_0$, $r_0\sim 0.1$ $nm$, the second system is a cube of the same size which is formed from $1000$ small cubes with $20$ electrons in each. These small cubes are assumed to be in thermal contact i.e. they are kept at a common temperature but the mutual particle exchange is impossible. The top panel of Fig.~\ref{Rubik} shows that at low temperatures the heat capacity of small systems, $N<10^{\,3}$, is practically equal to zero and only behind $N\sim 10^{\,4}$ $C$ gains the Sommerfeld (linear in $T$) values $C_{lin}$.  Therefore the canonical heat capacity corresponding to $10^{\,3}$ small cubes within the temperature interval $0\leq T/\varepsilon_F\leq 0.04$ (the bottom panel of Fig.~\ref{Rubik}) remains considerably smaller than $C$ of the ``large'' cube with $N=2\cdot 10^{\,4}$ electrons. Thus, regulating the size of mesoscopic grains in a sample gives a possibility to profoundly affect the thermodynamic characteristics of this sample at low temperatures. The bottom panel of Fig.~\ref{Rubik} indicates once more that as a rule the $EGCE$ method overestimates the values of $C$ at low temperatures.

In spherical cavities heating uncovers periodic structures oscillating with $\Delta N\sim N ^{\,1/2}$. These oscillations are closely connected with distribution of shells within supershells. Such bunches of shells can be occupied by large quantity of particles $\sim N ^{\,1/2}$, ~\cite{bohr}, accordingly the energy intervals between such level bunches $\Delta\varepsilon\sim\varepsilon_F/N^{\,1/2}$.

As mentioned above (see Sec.4) the values of $C$ can be interpreted as the averaged values of the level density i.e. studying $C$ v.s. $N$ one can get the averaged $\rho$ as a function of $N$. Theoretically these quantities were calculated by means of averaging $\rho(\varepsilon)$ either with Lorentzians~\cite{nishioka} or with Gaussians~\cite{brack2}, in both cases the widths of these averaging functions were temperature independent and of order of $0.5T_{sm}$. It is evident that at such temperature the result of averaging weakly depends on the type of averaging functions.

In the frame of the $EGCE$ formalism each oscillating functions ($n$-mode) in $\rho(\varepsilon)$, i.e. $\cos(2\pi\varepsilon/\tau_n+\phi)$ for
oscillators or $\cos(4\pi\sqrt{\varepsilon_F\varepsilon}/\tau_k+\widetilde{\phi})$ for cavities, $\phi$ and $\widetilde{\phi}$ being independent phases, after averaging with $\varphi(\varepsilon)$-functions ( for the sake of simplicity we suppose here that $\lambda+\beta\partial\lambda/\partial\beta\simeq\lambda$) enters into $C$ for oscillators or cavities respectively as $cos(2\pi\lambda/\tau_n+\phi)$ or $cos(4\pi\lambda/\tau_n+\widetilde{\phi})$. Thus the $N$-oscillations of $C$ in the $EGCE$ approach (see Fig.~\ref{C3DoscCompTeF01}) is caused by variations of values of $\lambda/\tau_n$. For those temperatures, when the oscillations could be observable,
$\lambda\simeq\varepsilon_F$ that for large $3D$-systems ($N\gg 1$) results in $\lambda/\tau_n\sim N^{1/3}$. Therefore in all figures the $N$-oscillations  in $C$ for $3D$-systems are presented as functions of $N^{1/3}$.

\begin{figure}
\scalebox{0.7}{\includegraphics{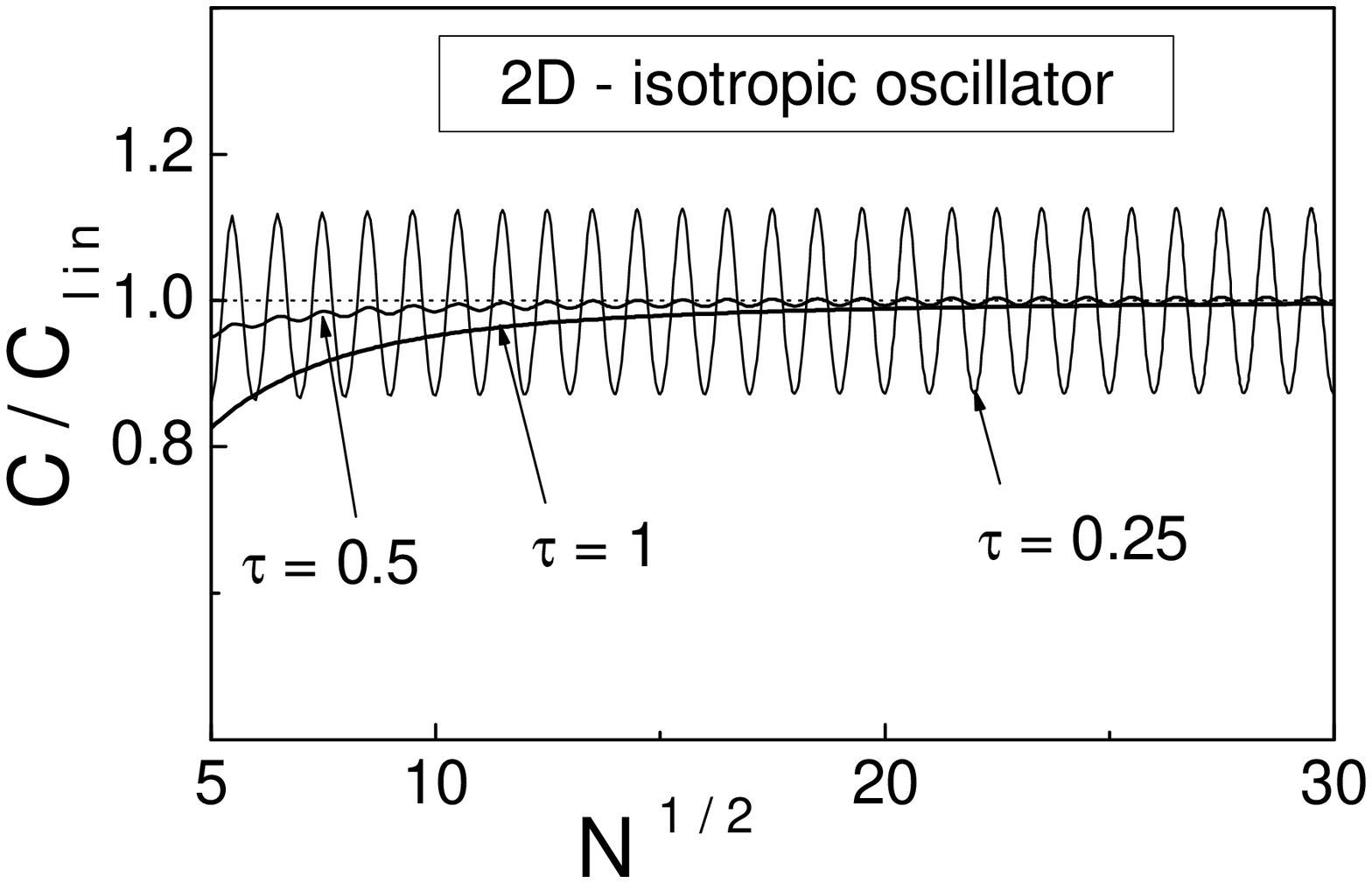}}
\caption{\label{2Dosc} $N$-oscillations of the heat capacity ($C/C_{lin}$) in $2D$-oscillators at different reduced temperatures $\tau=(T/\varepsilon_F)N^{1/2}$.}
\end{figure}

\begin{figure}
\scalebox{0.5}{\includegraphics{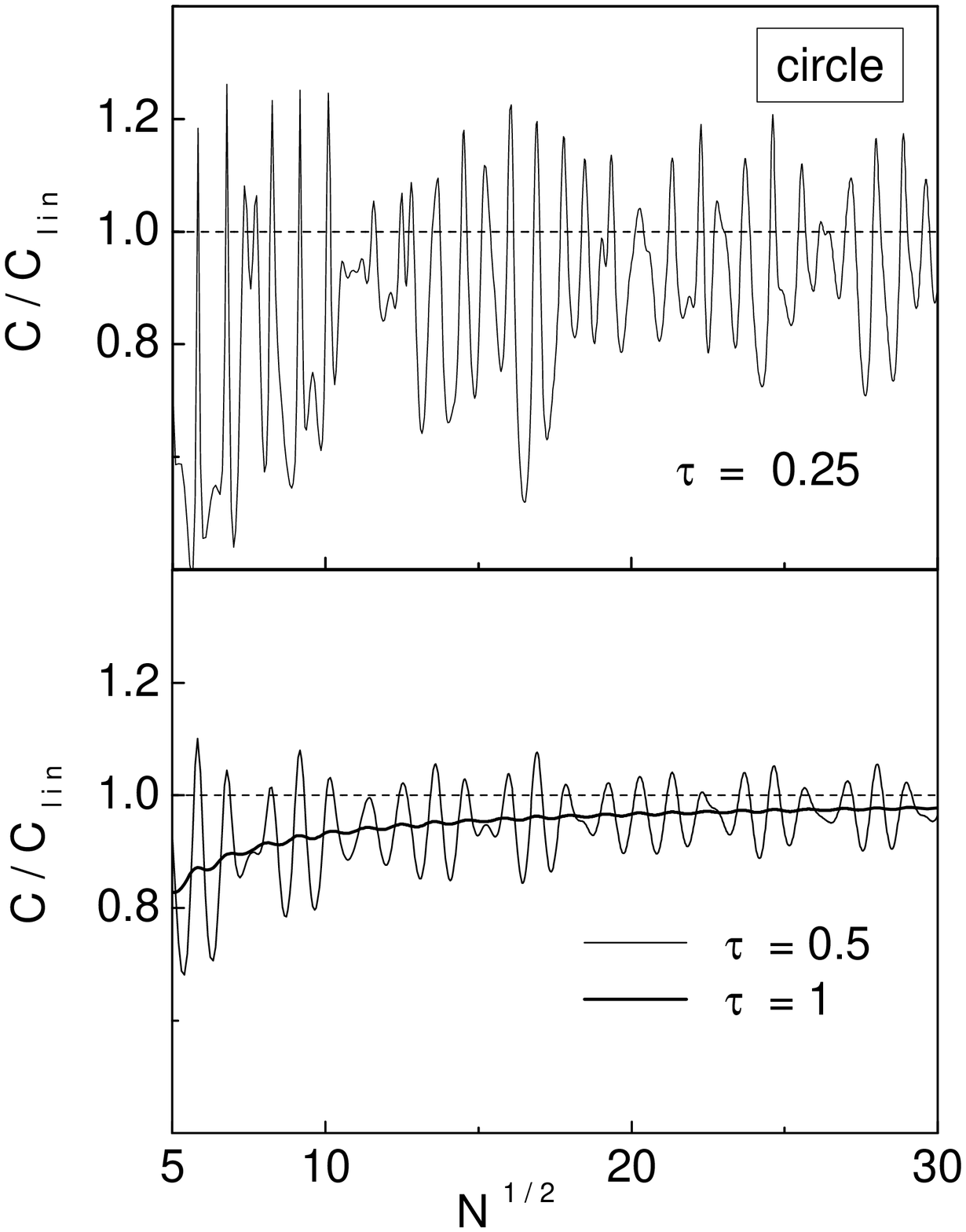}}
\caption{\label{CircleOsc} $N$- oscillations of the heat capacity ($C/C_{lin}$) vs $N^{1/2}$ in circles at three reduced temperatures $\tau=(T/\varepsilon_F)N^{1/2}$.}
\end{figure}

\begin{figure}
\scalebox{0.5}{\includegraphics{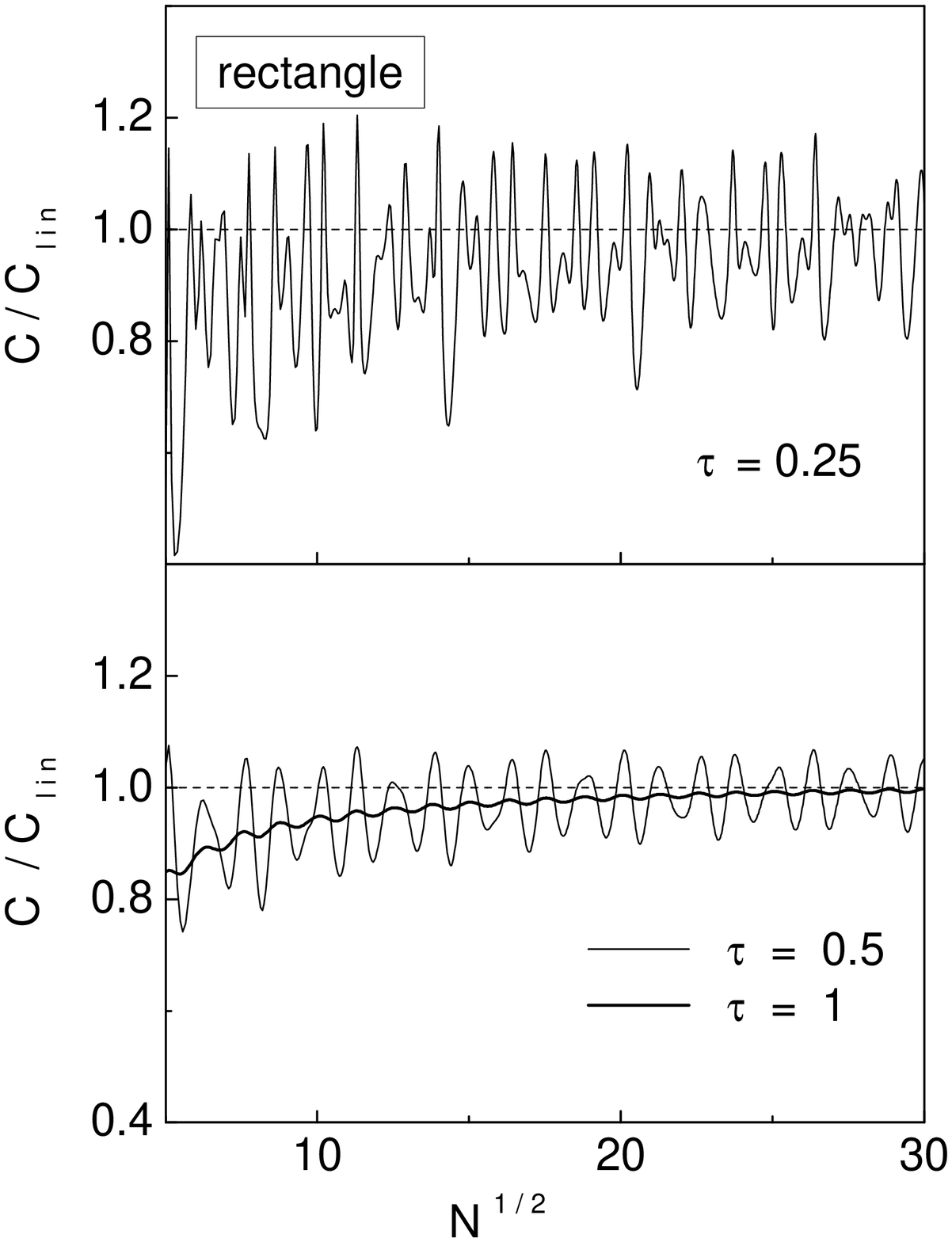}}
\caption{\label{RectOsc} $N$- oscillations of the heat capacity ($C/C_{lin}$) vs $N^{1/2}$ in rectangles ($L_x:L_y=1:e/\pi$) at three reduced temperatures $\tau=(T/\varepsilon_F)N^{1/2}$.}
\end{figure}

All arguments we have used for $3D$-systems can be repeated for $2D$-systems. The difference is only in the $N$-dependence of oscillating periods $\tau_0$ and $\tau_n$. For $2D$-systems
\begin{equation}\label{tau0}
\tau_0\sim\frac{\varepsilon_F}{\sqrt{N}};\;\;\;\;T_{sm}\sim\tau_0,
\end{equation}
and the $N$-oscillations have to be functions of $N^{1/2}$: These conclusions are confirmed by examples of the $N$-oscillations of $C$ in $2D$ isotropic oscillators ( Fig.~\ref{2Dosc}) and in $2D$-systems with rectangular potential such as circles (Fig.~\ref{CircleOsc}) and rectangles (Fig.~\ref{RectOsc}). The suppershell pattern in circles is pronounced not so distinctly as in spheres and has a shorter beating period so that at $\tau=(T/\varepsilon_F)N^{1/2}>0.5$ the $N$-oscillations in circles and rectangles are practically identical. For circles the oscillating structures in the level densities are considered in Ref.~\cite{tatievski}.
\begin{figure}
\scalebox{0.7}{\includegraphics{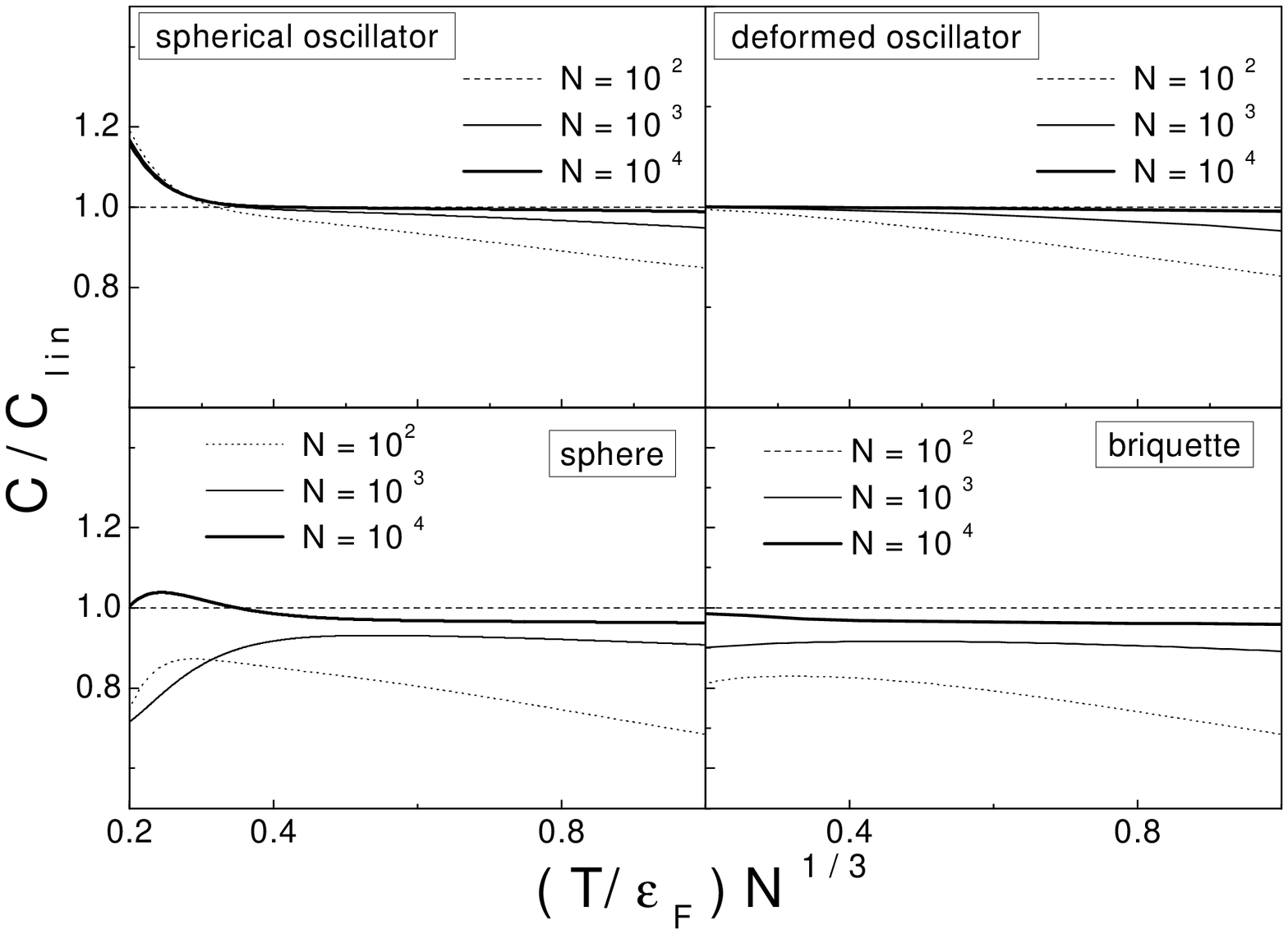}}
\caption{\label{CClinTN13eF} The reduced  ($C/C_{lin}$) heat capacity vs $\tau=(T/\varepsilon_F)N^{1/3}$ (in the temperature region where $C$ is $T$ linear  for large $N$)  for $3D$- spherical (a) and deformed (b) oscillators and for spherical (c) and briquette (d) systems in rectangular cavities   at different $N$.}
\end{figure}

This process which embraces  beginnings of the $N$-oscillations in $C$, appearance of periodic structures and their subsequent damping with $T$ is in parallel to the de Haas-van Alphen effect in mesoscopic systems considered in our work, Ref.~\cite{kuzmenko2}, where variations of $C$ and magnetic susceptibility  in a linearly increasing magnetic field $H$ were demonstrated at different temperatures. The analytical aspects of the analogy between $C$ vs $N$ and $C$ vs $H^{-1}$ were considered in Ref.~\cite{toms}.

At heating beyond $T_{sm}$ the heat capacity is determined only by $\rho_0(\varepsilon)$, the smooth level density, and the difference in occupation numbers does not affect the $N$-dependence of $C$. This at least for very large systems, $N\gg 1$, has to result in the $T$-linear regime of the heat capacity variations. In fact, see Fig.~\ref{CClinTN13eF}, for oscillators with $N>10^{\,4}$ the reduced heat capacity $C/C_{lin}$ at $T\sim T_{sm}$ is approximately equal to $1$  whereas for lower values of $N$ in oscillators and for all $N$ including $10^{\,4}$ in cavities $C/C_{lin}<1$.

For small $N$ ($N<10^{\,3}$) temperatures of order of $T_{sm}$ turn out to be so high, $T_{sm}(N=10^2)\sim\varepsilon_F/5$ that the condition determining the $T$ linear regime for $C$ ($T\ll\varepsilon_F$) is not implemented. Practically after and near $T_{sm}$ such few particle systems start going over to the classic limit of $C$ (this transition regime we called in Sec.2 as the quasilinear one). In systems confined in cavities with $N>10^{\,3}$ the heat capacity in a temperature interval $T>T_{sm}$ displays the $T$-linear dependence (Fig.~\ref{CClinTN13eF}). However for a fixed shape of the system the factor of proportionality in $C/N$ varies with the particle number that testifies to the more complicated $N$-dependence of $\rho_0(\varepsilon)$ as compared with the case $N\gg 1$.

\section{Deceleration of increasing $c$ in temperature range $T_{sm}<T<\varepsilon_F$}

On  going out towards the linear dependence on $T$ the heat capacity slowly evolves to the classic Boltzmann-Maxwell limit. Consequently there exists a temperature range ($T>T_{sm}$) within which the growth of $C$ decelerates to attain to the saturation. To find the value of $T$ giving visible deviations from the linear low at $N\gg 1$ the smooth level density $\rho_0(\varepsilon)$ can be represented in the form:
\begin{equation}\label{rhoegamma}
\rho_0(\varepsilon)=\gamma\frac{N}{\varepsilon_F}\left (\frac{\varepsilon}{\varepsilon_F}\right )^{\gamma-1},
\end{equation}
then the integral for $C$, Eq.~(\ref{C}), with $\rho_0(\varepsilon)$ instead of $\rho_{ex}(\varepsilon)$ can be calculated by the standard way up to terms $(\beta\varepsilon_F)^{-2}$ and $(\beta\varepsilon_F)^2\exp(-\beta\varepsilon_F)$

\begin{equation}\label{Ctotal}
C/k_B=\gamma\frac{\pi^2}{3}N\frac{T}{\varepsilon_F}\left \{ 1-\frac{(\gamma-1)(9-2\gamma)\pi^2}{10(\beta\varepsilon_F)^2}
-\frac{3(\gamma-2)(\gamma-3)}{2\pi^2}(\beta\varepsilon_F)^2e^{-\beta\varepsilon_F}\right\}.
\end{equation}

For few particle systems Eq.~(\ref{Ctotal})can be used to obtain rough estimates of such temperature. The origin of the last term in Eq.~(\ref{Ctotal}) is the finiteness of the value of $\varepsilon_F/T$ that is usually supposed to be infinitely large. However it is not the case if $(\beta\varepsilon_F)^{-1}$ is of order of several tenths. Eq.~(\ref{Ctotal}) shows that $(20\div 30)\%$ deviations from linearity are observed at $(\beta\varepsilon_F)^{-1}\sim 0.3$ for $\gamma=3/2$ ( $3D$-cavities) and at $(\beta\varepsilon_F)^{-1}\sim 0.2$ for $\gamma=3$ ($3D$-oscillators). Absolute values of this temperature corresponding to these deviations are material dependent. So for metal clusters it is $\sim 10^{\,3}\div10^{\,4}$ $K$ while for heterostructures $\sim 100$ $K$.

\begin{figure}
\scalebox{0.7}{\includegraphics{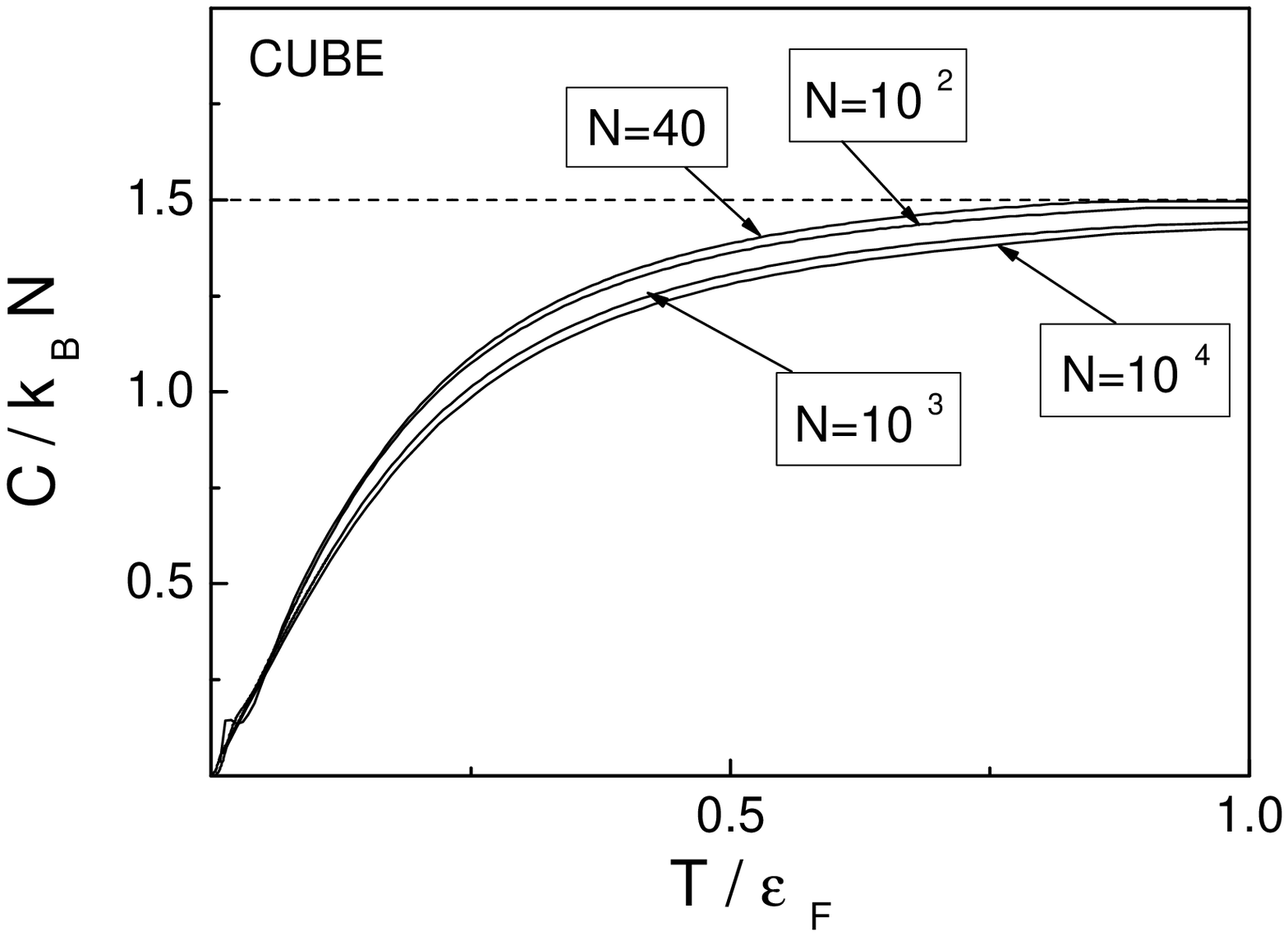}}
\caption{\label{CubeManyN} The specific heat in a cube ($\gamma =1.5$) for  different $N$ at high temperatures.}
\end{figure}

Fig.~\ref{CubeManyN} displaying the specific heat in cavities demonstrates the dependence of the specific heat on $N$. This is the obvious manifestation of the complicated $N$-dependence of the smooth level density $\rho_0(\varepsilon)$ that was mentioned in the end of the previous section.

For $3D$-cavities, $\gamma=3/2$, $\rho_0(\varepsilon)$ consists of the volume ($V$), surface ($S$) and linear terms~\cite{balian}
\begin{equation}\label{rho3terms}
\rho_0(\varepsilon)=\frac{V}{2\pi^2}\left (\frac{\hbar^2}{2m}\right )^{-3/2}\sqrt\varepsilon-
\frac{S}{4\pi}\left (\frac{\hbar^2}{2m}\right )^{-1} + \rho_{lin}\left (\frac{\hbar^2}{2m}\right )^{1/2}\frac{1}{\sqrt\varepsilon}.
\end{equation}
The explicit dependence of $\rho_0(\varepsilon)$ on the sizes of the system can be avoided due to the relationship between $\varepsilon_F$ and  a parameter of length inherent in the system, Eq.~(\ref{eFRL}). Such parameter for spheres is the radius $R$, for cubes the lateral length $L$. The same quantity $L$ is suitable for briquette  if $L=(L_xL_yL_z)^{1/3}$
\begin{equation}\label{eFRL}
\varepsilon_F=\frac{\hbar^2}{2mR^2}X_F^2\;\;\;\; or \;\;\;\; \frac{\hbar^2}{2mL^2}X_F^2,
\end{equation}
where $X_F$ for spheres is a Bessel function root, for briquettes
\begin{eqnarray}\label{XF}
X^2_F =\pi^2\left ( n_x^2/\alpha_x^2 + n_y^2/\alpha_y^2 +n_z^2/\alpha_z^2\right ),\\
\alpha_x:\alpha_y:\alpha_z=L_x:L_y:L_z,\;\;\;  \alpha_x\alpha_y\alpha_z=1, \nonumber
\end{eqnarray}
For cubes $\alpha_x=\alpha_y=\alpha_z=1$.

Then, the replacement of  $\hbar^2 /2mR^2$ or $\hbar^2/2mL^2$ by $\varepsilon_F/X_F^2$ reduces Eq.~(\ref{rho3terms}) to
\begin{equation}\label{rhoegamma1}
\rho_0(\varepsilon)=\frac{N}{\varepsilon_F}\sum_{\gamma=3/2;1;1/2}\gamma a_{\gamma}\left (\frac{\varepsilon}{\varepsilon_F}\right )^{\gamma-1}.
\end{equation}
Coefficients $a_{\gamma}$ are straightforwardly found by comparison of Eqs.~(\ref{rho3terms}),(\ref{eFRL}) and Eq.~(\ref{rhoegamma1}):
\begin{equation}\label{agamma}
a_{\gamma}\sim X_F^{2\gamma}/N.
\end{equation}
Calculations of $C$ with $\rho_0(\varepsilon)$, Eq.~(\ref{rhoegamma1}), lead to the result similar to Eq.~(\ref{Ctotal}), but the total factor determining the linearity in $T$ and factors before $(\beta\varepsilon_F)^{-2}$ and  $(\beta\varepsilon_F)^{2}\exp(\beta\varepsilon_F)$ gain a more complicated dependence on $N$ as compared with Eq.~(\ref{Ctotal}). In particular instead of $\gamma=3/2$ (the first factor in  Eq.~(\ref{Ctotal})) now we have $\sum_{\gamma}\gamma a_{\gamma} \left (\frac{\lambda_0}{\varepsilon_F}\right )^{\gamma-1}$. (The quantity $\lambda_0$ is defined by the equation $\sum_{\gamma}a_{\gamma}\left (\frac{\lambda_0}{\varepsilon_F}\right )^{\gamma}=1$).
For small $N$ ($<10^3$) this factor $>3/2$, that is caused mainly by coefficients $a_{3/2}$, Eq.~(\ref{agamma}), which for small $N$ is more than $1$. For $N\gg 1$ the dependence $X_F$ on $N$ can be established by using the integral independing of  $\varepsilon_F$

\begin{equation}\label{rhoagamma}
\frac{1}{N}\int_0^{\varepsilon_F}\rho(\varepsilon)d\varepsilon=\sum_{\gamma}a_{\gamma}.
\end{equation}
For large values of $N$ Eq.~(\ref{rhoagamma}) is practically equal to $1$. The surface and linear term in $\rho(\varepsilon)$ at $N>10^3$ are unessential and so from Eq.~(\ref{rhoagamma}) it follows
\begin{equation}\label{X3F}
X_F^3=9\pi N/4\;\;\;(sphere)\;\;\;or\;\; 3\pi^2 N(cube\;\;\;or\;\;\;briquette) .
\end{equation}
However at small $N$ the negative surface term $(\sim X^2_F)$ plays the important role, raising the values of $X_F$ as compared with Eq.~(\ref{X3F}). This explains exceeding of $C/N$ for small $N$ over large ones at the same temperature in Figs.~\ref{CubeManyN}.

The values of $C/N$ for small N can be affected even by the linear term in $\rho_0(\varepsilon)$. The surface and linear terms in $C$ have different dependence on shape parameters of briquettes, Eq.~(\ref{XF}): the surface term $\sim\sum\alpha_i^{-1}$ and linear term  $\sim\sum\alpha_i$. These sums can be essentially distinguished for strongly stretched and flattened shapes. That can give rise to difference in $C$ at small $N$.

\begin{figure}
\scalebox{0.5}{\includegraphics{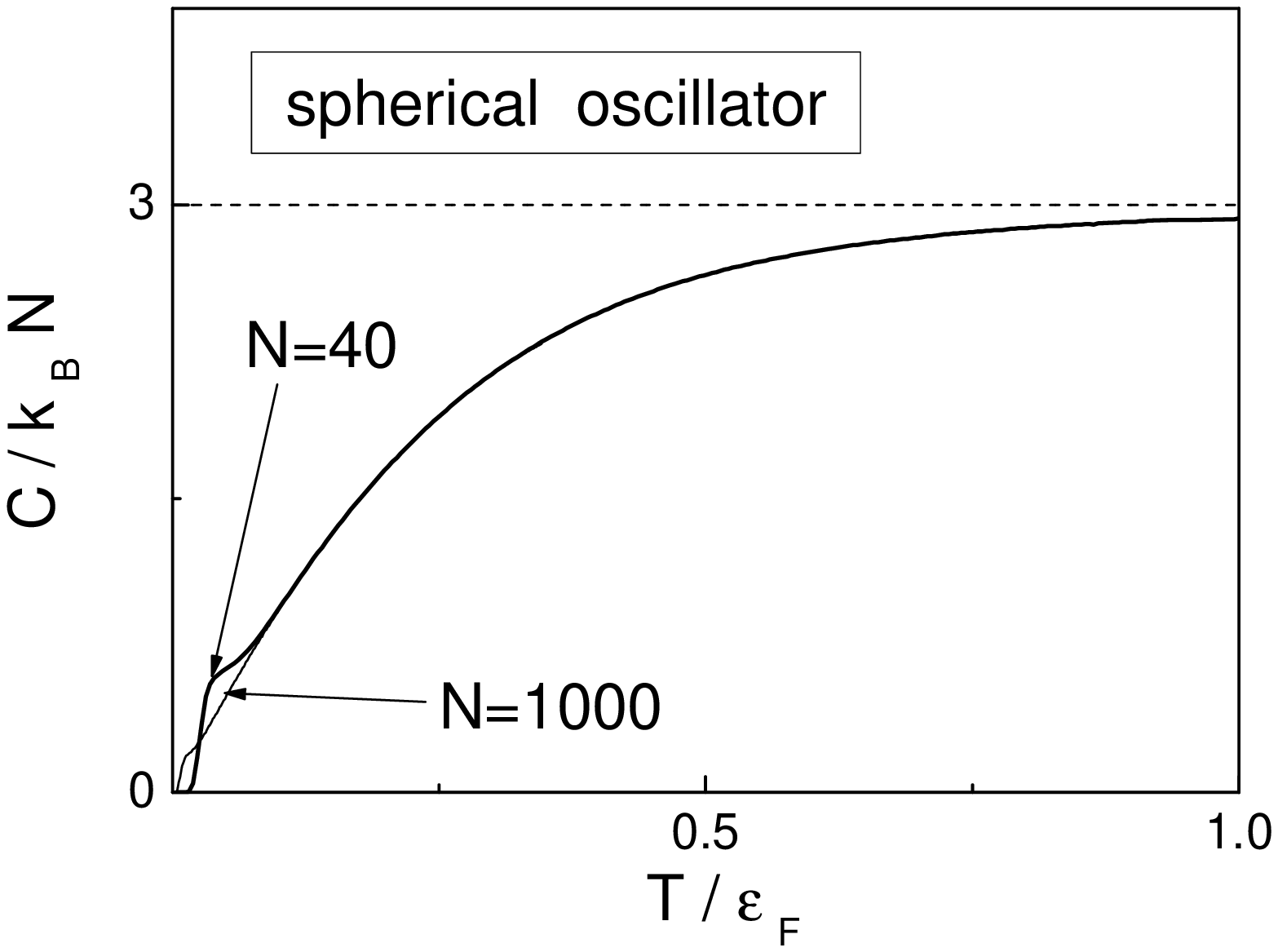}}
\caption{\label{OscClasLim} The specific heat in a spherical oscillator ($\gamma =3$) for  $N=40$ and $1000$ at high temperatures.}
\end{figure}

Fig.~\ref{OscClasLim}, where $C/N$ for a spherical oscillator are displayed, shows that $N$-dependence of $C/N$ for oscillators at $T\sim\varepsilon_F$ are practically absent since for this type of confinement, $\gamma =3$, the additional term in $\rho(\varepsilon)$ is much less than the first ($\sim N^{-2/3}$)
\begin{equation}\label{rhoosc1}
\rho_0(\varepsilon)=(\omega_x\omega_y\omega_z)^{-1}\left [\varepsilon^2-\frac{1}{12}(\omega_x^2+ \omega_y^2+\omega_z^2)\right ],\;\;\;\omega_i\sim N^{-1/3}.
\end{equation}
Only for small $N$ and strongly deformed oscillators the second term in Eq.~(\ref{rhoosc1}) could play some role.

Thus, the variations $C/N$ vs $T$ at $T>T_{sm}$ indicate that the factor determining the linearity in $T$ can get the $N$-dependence caused by the type of confinement. Probably realistic potentials are something middle between rectangular and oscillator potentials. Therefore studying the specific heat could give additional information concerning the type of confinement in a fermion system under consideration.

Variations of $\varepsilon_F$ vs $N$ in fermion mesoscopic systems could be predicted by a theory ascertaining connection of $N$ with linear sizes of the system. Here we will consider the jellium model postulating  that the linear sizes are proportional to $N^{\,1/3}$
\begin{equation}\label{RLN13}
R=r_0 N^{1/3} (sphere),\;\;\; L=l_0N^{1/3} (briquette),
\end{equation}
$r_0$, $l_0$ being independent of $N$.

Then, applying Eqs.~(\ref{eFRL}),~(\ref{X3F}),~(\ref{RLN13}) one obtains
\begin{eqnarray}\label{eFNsphere}
\frac{\varepsilon_F(N)}{\varepsilon_F^{(bulk)}}=\frac{X^2}{N^{2/3}}\left (\frac{4}{9\pi}\right )^{2/3} (sphere)\\
\frac{\varepsilon_F(N)}{\varepsilon_F^{(bulk)}}=\frac{X^2}{N^{2/3}}\left (\frac{1}{3\pi^2}\right )^{2/3} (briquette),\label{eFNbrik}
\end{eqnarray}
where  $\varepsilon_F^{(bulk)}$ is the bulk Fermi energy.
\begin{figure}
\scalebox{0.5}{\includegraphics{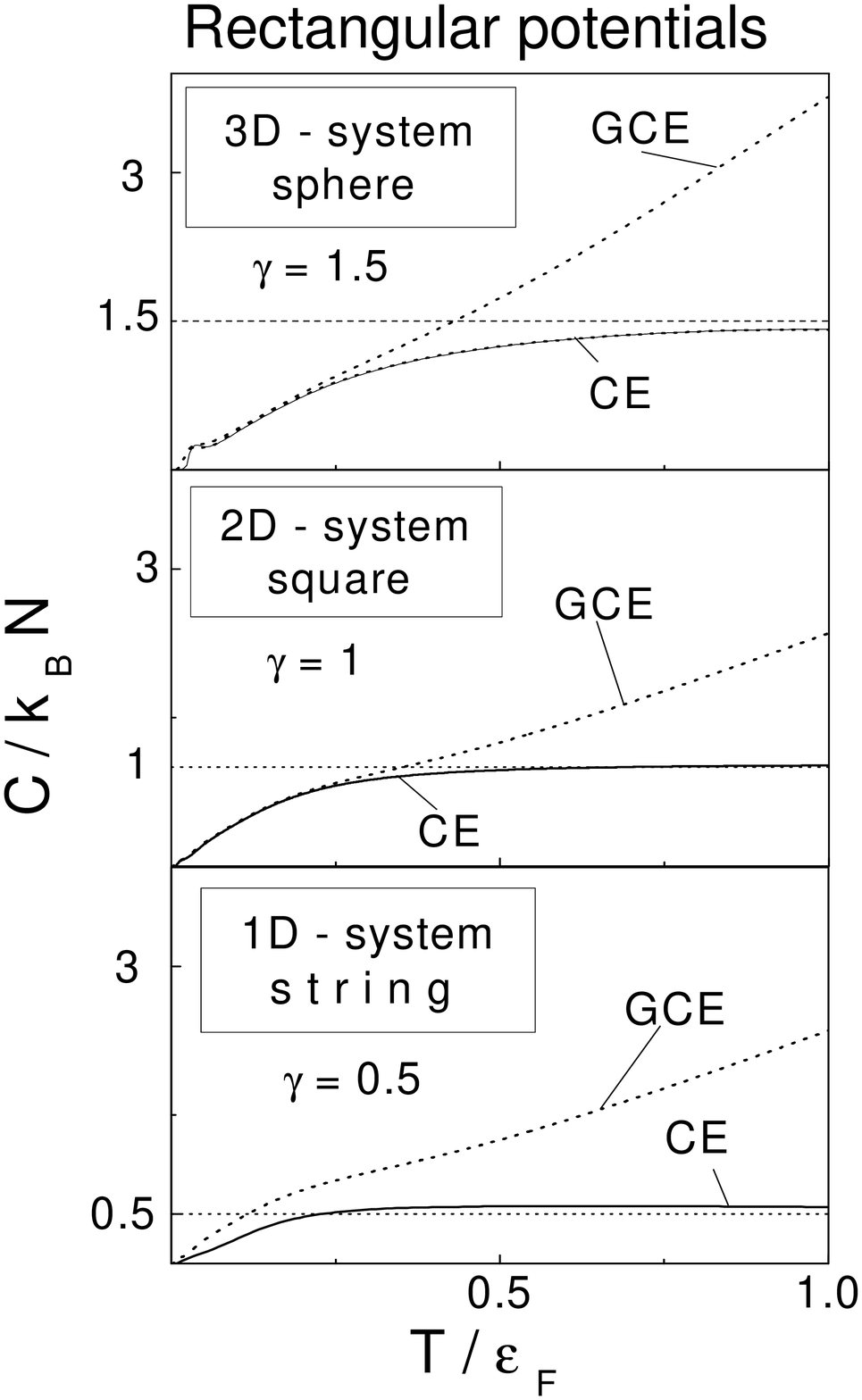}}
\caption{\label{RectPotGamma} Comparison of the values of $C$ calculated by three methods: $CE$,  $EGCE$ and $GCE$ (see Sec.2). Calculations are performed for systems with $40$ fermions confined in rectangular potentials of different spatial dimensions. On the scale of the figure $C(EGCE)$ coincides with $C(CE)$ at all temperatures. At $T>0.2\varepsilon_F$ the $GCE$ method overestimates the values of $C$ and gives wrong results at $T\longrightarrow\infty$.}
\end{figure}
\begin{figure}
\scalebox{0.5}{\includegraphics{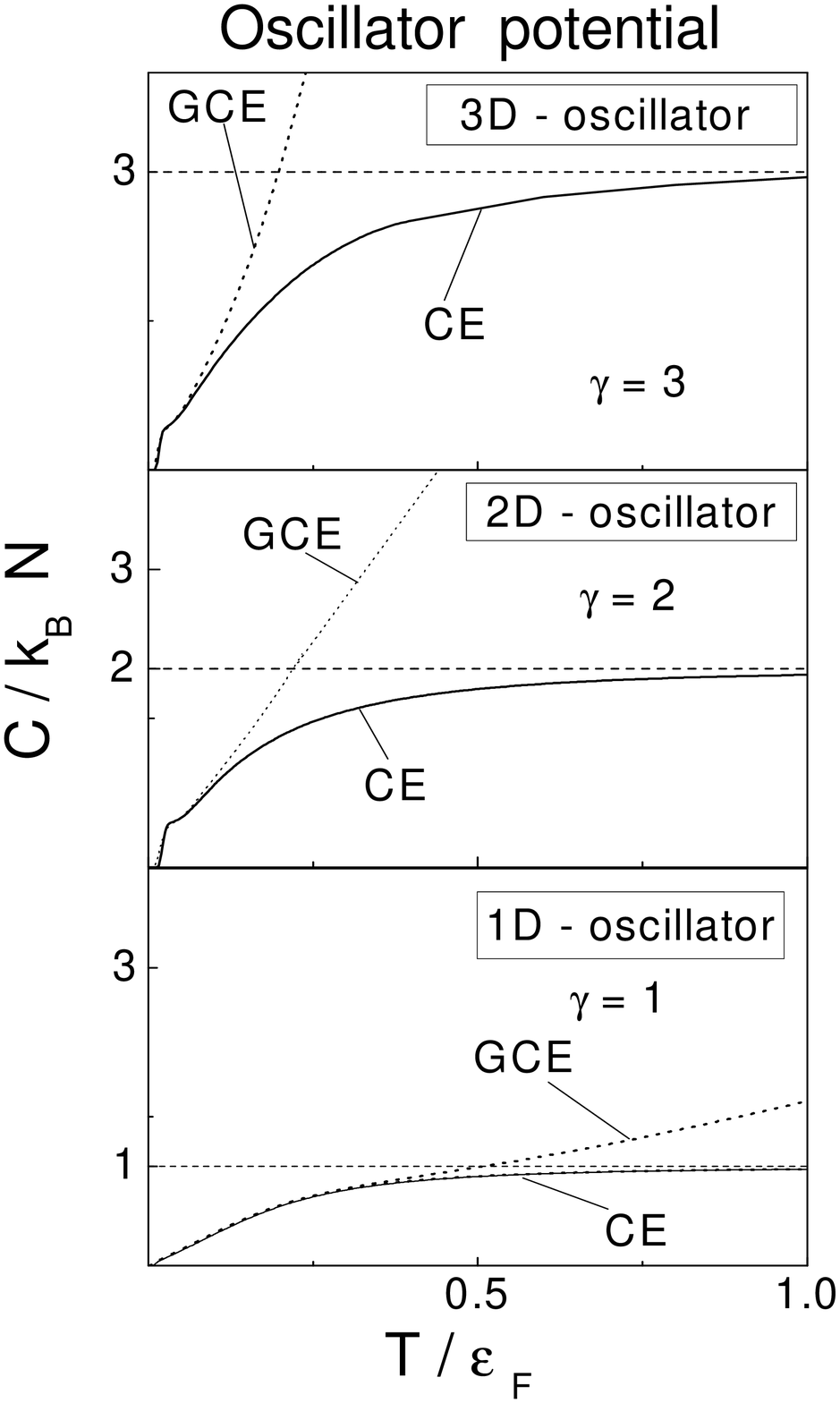}}
\caption{\label{OscPotGamma} The same as in Fig.~\ref{RectPotGamma} but for the isotropic oscillators. On the scale of the figure $C(EGCE)$ coincides with $C(CE)$ at all temperatures.}
\end{figure}
\begin{figure}
\scalebox{0.5}{\includegraphics{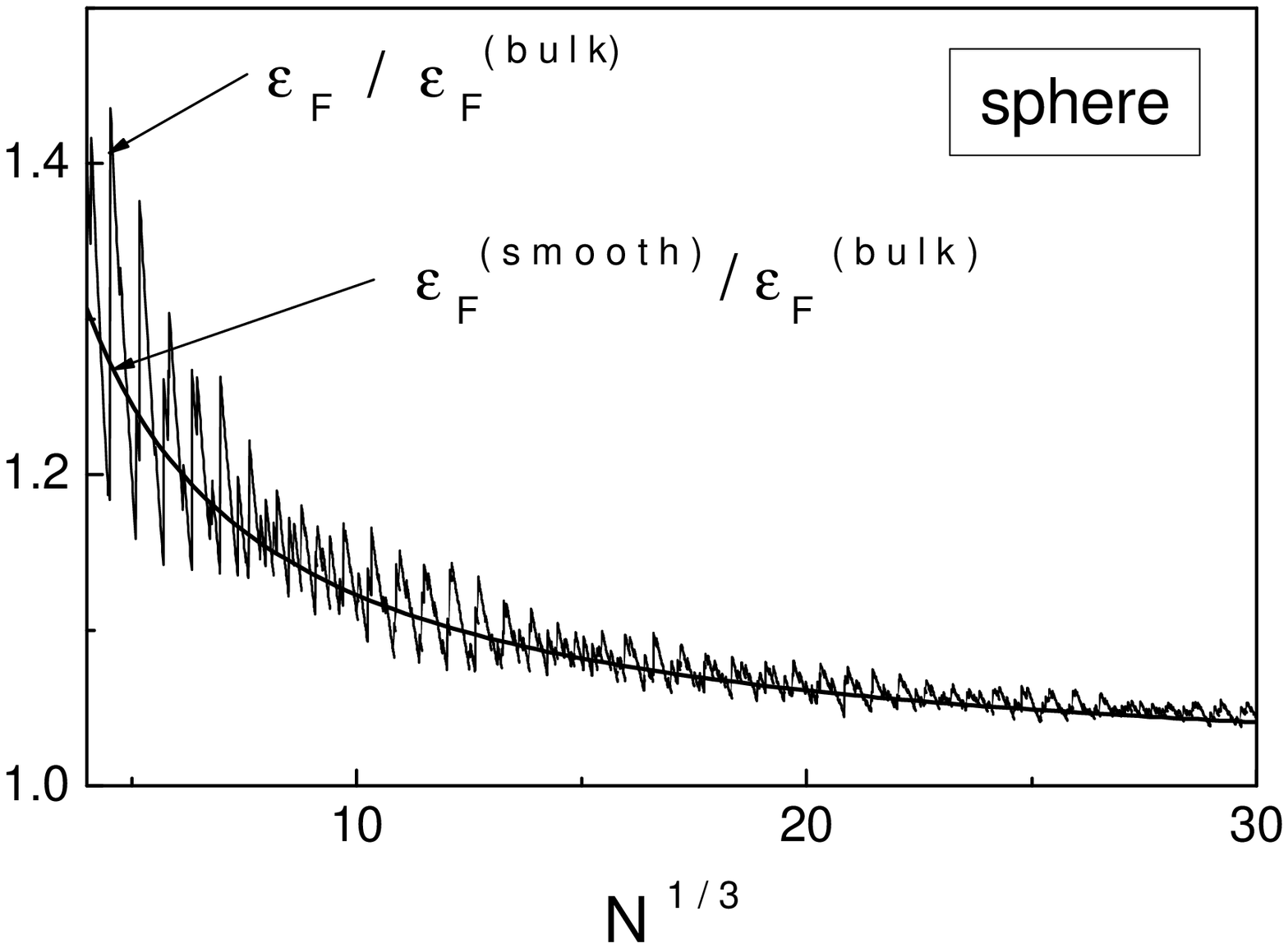}}
\caption{\label{sphEfermi}The Fermi energy $\varepsilon_F$ (solid line) in spheres vs $N$ in the jellium model, Eq.~(\ref{eFNsphere}), where $r_0$ is
material dependent parameter independent of $N$, ($R=r_0N^{1/3}$,  $r_0$ is the radius of a sphere containing one particle). $\varepsilon_F^{(smooth)}$ is  the  smooth Fermi energy, Eq.~(\ref{spheFsmooth}).}
\end{figure}

The results of calculations of $\varepsilon_F(N)$, Eq.~(\ref{eFNsphere}), for spherical systems are displayed in Fig.~\ref{sphEfermi}. In the same figure the smooth curve is plotted. It is obtained by replacing the exact values of $X_F$ by approximate ones found by using Eq.~(\ref{rhoagamma}) (assuming that $\sum a_{\gamma}=1$). This curve allowing for only the surface term (see also ~\cite{robnik}) in $\rho_0$, Eq.~(\ref{rho3terms}), gives the following smooth dependence $\varepsilon_F^{(smooth)}$ on $N$:
\begin{eqnarray}\label{spheFsmooth}
\frac{\varepsilon_F^{(smooth)}}{\varepsilon_F^{(bulk)}}\simeq 1+\left (\frac{3\pi^2}{16N}\right )^{1/3}\;\;\; (sphere)\\
\frac{\varepsilon_F^{(smooth)}}{\varepsilon_F^{(bulk)}}\simeq 1+\left (\frac{\pi}{3N}\right )^{1/3}(\sum_i\alpha^{-1})^{1/2}\;\;\; (briquette),\label{brikeFsmooth}
\end{eqnarray}
the definition of $\alpha_i$ is given by Eq.~(\ref{agamma}). Fig.~\ref{sphEfermi} and Eqs.~(\ref{spheFsmooth}), ~(\ref{brikeFsmooth}) show that in the model in which the parameter of spatial density ($r_0$ or $l_0$), Eq.~(\ref{RLN13}), is independent of $N$ the energy $\varepsilon_F^{(N)}$ oscillates with $N$ around $\varepsilon_F^{(smooth)}$ and even for small particle numbers ($N\sim 50$) $\varepsilon_F^{(N)}$ deviates from $\varepsilon_F^{(bulk)}$ not more than by $40\%$.

\section{The classic limit of $C$}

Though in the temperature range $T_{sm}<T<\varepsilon_F$ the specific heat ($C/N$) reveals some dependence on $N$, in the limit of superhigh temperature $T/\varepsilon_F\gg 1$ this dependence completely dies out and irrespectively of the shape and particle numbers $C/N$ trends to the classic Boltzmann-Maxwell limit
\begin{equation}\label{gammalimit}
C/k_BN=\gamma
\end{equation}
(remind that $\gamma=D/2$ for cavities and $\gamma=D$ for oscillators).
The comparison of $C$ obtained by three methods ($CE$, $GCE$and $EGCE$, see Sec.2) in the high temperature region up to $\varepsilon_F$ is given in Figs.~\ref{RectPotGamma},~\ref{OscPotGamma}. It is obviously that at high temperatures and at $T\longrightarrow\infty$ for all systems $EGCE$ gives practically the same results as $CE$ and both methods lead to the correct classic limit while $GCE$ gives wrong results for $C$ at $T>0.2\varepsilon_F$.
Consider the behavior of the specific heat  in the limit of superhigh temperature, $T/\varepsilon_F\gg 1$, in more details by using the $EGCE$ formalism and begin with the calculation of $N$ to find $\beta\partial\lambda/\partial\beta$
\begin{equation}\label{NhighT}
N(T\gg T_{sm})=\int_0^{\infty}\rho_0(\varepsilon)e^{\beta (\varepsilon - \lambda)}d\varepsilon
\end{equation}
The simple form of Eq.~(\ref{NhighT}) takes place since at $T\gg\varepsilon_F$ the chemical potential becomes negative $\lambda <0$ ( the energy scale is chosen so that all single particle level energies are positive). Taking advantage of the expansion of $\rho_0(\varepsilon)$ in powers of $\varepsilon$ one can transform Eq.~(\ref{NhighT}) into the following
\begin{eqnarray}
N=N\sum_{\gamma}\gamma a_{\gamma}e^{\beta\lambda}\varepsilon_F^{-\gamma}\int_0^{\infty}\varepsilon^{\gamma -1}e^{-\beta\varepsilon}d\varepsilon = \nonumber \\
=N\sum_{\gamma}\gamma a_{\gamma}e^{\beta\lambda}(\beta\varepsilon_F)^{-\gamma}\Gamma (\gamma )\label{Ngamma}.
\end{eqnarray}
Eq.~(\ref{Ngamma}) indicates that the more is the temperature the less terms with $\gamma <\gamma_{max}$ as compared with the first term with $\gamma =\gamma_{max}$. Below index ``max'' will be omitted. In this approximation Eq.~(\ref{Ngamma}) defines $\beta\lambda$ and $\beta^2\partial\lambda/\partial\beta$:
\begin{eqnarray}\label{betalambda1}
\beta\lambda =-\left [\gamma\ln\frac{T}{\varepsilon_F}+ \ln a_{\gamma}+\ln\Gamma (\gamma+1)\right ] ;\\
\beta\left (\lambda + \beta\frac{\partial\lambda}{\partial\beta}\right )=e^{\beta\lambda}\gamma a_{\gamma}\Gamma (\gamma+1)\left(\frac{T}{\varepsilon_F}\right )^{\gamma}=\gamma \label{betalambda2}.
\end{eqnarray}

Comparison of Eq.~(\ref{betalambda1}) and Eq.~(\ref{betalambda2}) emphasizes the importance of the term $\partial\lambda/\partial\beta$ because only taking into account this term makes the quantity $\beta (\lambda + \beta\partial\lambda/\partial\beta)$ independent of $T$.

Also as in Eq.~(\ref{Ngamma}) the calculations of $C$ is performed with the only term in $\rho(\varepsilon)$ with maximum $\gamma$.
\begin{equation}\label{Cgamma1}
C/k_B=N\gamma a_{\gamma} e^{\beta\lambda}\varepsilon_F^{-\gamma}\int_0^{\infty}(x-\gamma)^2\varepsilon^{\gamma -1}e^{-x}d\varepsilon
\end{equation}

In this equation $x=\beta\varepsilon$ and $\beta(\lambda + \beta\frac{\partial\lambda}{\partial\beta})$ is replaced by $\gamma$ in accordance with Eq.~(\ref{Ngamma}).
\begin{equation}\label{Cgamma2}
C/k_B=N\gamma a_{\gamma} e^{\beta\lambda}(\beta\varepsilon_F)^{-\gamma}\Gamma(\gamma)\left [(\gamma+1)\gamma -2\gamma^2+\gamma^2\right ].
\end{equation}
Comparison of first factors ( before square brackets) with Eq.~(\ref{Ngamma}) gives the final result, Eq.~(\ref{gammalimit}). The same result can be obtain  in the canonical formalism starting with the canonical partition function $Z_N$. As shown in Ref.~\cite{kuzmenko} $Z_N$ is equal to $[N]$, a symmetrical polynomial of the power $N$ in variables $q_s=\exp(-\beta\varepsilon_s)$, $\varepsilon_s$ being the single-particle energy of state $s$. According to the known Warring formulae each symmetrical polynomial can be represented through power sums $S_n$
\begin{equation}\label{Sn1}
S_n=\sum_{s}d_sq_s^n=\int_0^{\infty}\rho_{ex}(\varepsilon)e^{n\beta\varepsilon}d\varepsilon,
\end{equation}
at high temperatures $\rho_{ex}(\varepsilon)\longrightarrow\rho_0(\varepsilon)$.
We write out only two first terms of this representation to show that the second term and all others disappear at $T\longrightarrow\infty$
\begin{equation}\label{Nbracket}
[N]=\frac{1}{N!}S_1^N-\frac{1}{(N-2)!2}S_2(S_1)^{N-2}+\ldots
\end{equation}

For superhigh temperatures:
\begin{eqnarray}\label{Sn2}
S_n=\frac{N}{n}a_{\gamma}(n\beta\varepsilon_F)^{-\gamma}\Gamma (\gamma),\\
S_2/(S_1)^2=(\beta\varepsilon_F)^{\gamma}/N2^{\gamma +1}\Gamma (\gamma)\mid_{T\longrightarrow\infty}\longrightarrow 0 \nonumber
\end{eqnarray}
Thus, to find $C$ in the high temperature limit it is sufficient to take only the first term in  Eq.~(\ref{Nbracket}) into account and use Eq.~(\ref{Sn2}) at $n=1$ that straightforwardly gives rise to the final result, Eq.~(\ref{gammalimit}):
\begin{equation}\label{Clog}
C/k_B=\beta^2\frac{\partial^2}{\partial\beta^2}\ln [N]=\beta^2\frac{\partial^2}{\partial\beta^2}\left \{N\ln S_1\right \}=\gamma N,
\end{equation}
since $\ln S_1=-\gamma\ln\beta$ $+$ terms independent of $\beta$.

\section{Conclusion}

The investigation of the heat capacity variations at heating from $T=0$ up to $T\sim\varepsilon_F$ allows us to infer what information can be obtained at measuring the fermion mesoscopic heat capacity.

At temperatures of order of or lesser than the average level spacing $\delta_{\,F}$ (i.e. $d_F\varepsilon_F/\gamma N$ where $d_F$ is the average level degeneration near the Fermi level) experiments make it possible, as it arises from Section $3$ and  $4$, to establish single -particle level structure  in the vicinity of $\varepsilon_F$. For this purpose two kinds of experiments could be performed on systems of the same type with a small variations in particle numbers. The first one is studying  low temperature local maxima or irregularities in the heat capacity temperature increase that gives the resonance temperature directly connected with $\varepsilon_{F+1}-\varepsilon_F$ (Sect.3) in each of such systems. The second kind of experiments on the same systems is measuring $C$ as a function of $N$ either at a fixed temperature smaller than $\delta_F$ or at temperatures of the same order but decreasing with $N$ ($\sim N^{-1/3}$ for $3D$-systems or $\sim N^{-1/2}$ for $2D$-ones). These experiments could give maxima  in $C$ for those values of $N$ that correspond to filled levels or shells. These experimental results ($C$ v.s. $T$ for adjacent $N$ or $C$ v.s. $N$ around some value $N_0$) can be used to reconstruct the energy level sequence near $\varepsilon_F$ corresponding to $N_0$.

As discussed in Sections $2$ and $4$ the heat capacity can be interpreted as the temperature averaged level density. Studying $C$ v.s. $N$ at $T\sim\delta_F$ gives too much details in the level density as a function of $N$. However at $T\sim 0.5T_{sm}$ ( behind $T_{sm}$ all level density oscillations discontinue, $T_{sm}\simeq\varepsilon_F N^{-1/3}$ for $3D$-systems and $T_{sm}\simeq\varepsilon_F N^{-1/2}$ for $2D$-ones) measuring $C$ uncovers the main periods of oscillations in $C$ and accordingly in the level density that can be applied to researching other mesoscopic phenomena  depending on the level density.

Besides the theoretical interest studying $C$ v.s. $N$ can find practical applications. For example, the heat conductivity is known to be proportional to the heat capacity and so variations of $C$ have to be taken into account if a nanogranule material is chosen as a thermoinsulator at very low temperatures when the electron heat capacity prevails over the lattice one. At such temperatures, as shown in Sec.4, the electron heat capacity of some mesoscopic systems, i.e. at some values of the particle number, takes practically zero values thereby such nanogranule  materials are also devoid of the heat conductivity.

This work is supported by the ISTC under grant Nr. 3492. The authors are much indebted to V.P. Chechev and R.B. Panin for the help in the work.

\end{document}